\documentclass[aps,showpacs,preprintnumbers,amsmath,amssymb]{revtex4}
\usepackage{dcolumn}
\usepackage[dvips]{epsfig}

\begin{document}
\baselineskip=0.8 cm
\title{\bf Strong gravitational lensing for the photons coupled to Weyl tensor in a Kerr black hole spacetime}

\author{Songbai Chen$^{1,2,3}$\footnote{csb3752@hunnu.edu.cn}, Shangyun Wang $^{1}$\footnote{shangyun\underline{~}wang@163.com}, Yang Huang $^{1}$\footnote{huangyang66@163.com} , Jiliang Jing$^{1,2,3}$
\footnote{jljing@hunnu.edu.cn}, Shiliang Wang$^{4}$}

\affiliation{$^{1}$Institute of Physics and Department of Physics, Hunan
Normal University,  Changsha, Hunan 410081, People's Republic of
China \\ $^{2}$Key Laboratory of Low Dimensional Quantum Structures \\
and Quantum Control of Ministry of Education, Hunan Normal
University, Changsha, Hunan 410081, People's Republic of China\\
$^{3}$Synergetic Innovation Center for Quantum Effects and Applications,
Hunan Normal University, Changsha, Hunan 410081, People's Republic
of China\\
$^{4}$School of Physics and Electronics,
Central South University, Changsha, Hunan 410083, China}

\begin{abstract}
\baselineskip=0.6 cm
\begin{center}
{\bf Abstract}
\end{center}

We present firstly equation of motion for the
photon coupled to Weyl tensor in a Kerr black hole spacetime
and then study further the corresponding strong
gravitational lensing.  We find that black hole
rotation makes propagation of the coupled photons more
 complicated, which brings about some new features for
physical quantities including the marginally
 circular photon orbit, the deflection angle,
 the observational gravitational lensing variables
 and the time delay between two relativistic images.
 There is a critical value of the coupling parameter
  for existence of the marginally circular photon orbit
   outside the event horizon, which depends on the rotation
    parameter of black hole and the polarization direction of
    photons. As the value of coupling parameter is near
     the critical value, we find that the marginally circular photon
     orbit for the retrograde photon increases with the rotation
     parameter, which modifies a common feature of the marginally
      circular photon orbit in a rotating black hole spacetime
      since it always decreases monotonously with the rotation
      parameter in the case without Weyl coupling. Combining
      with the supermassive central object in our Galaxy, we
      estimated the observables including time delays between
       the relativistic images in the strong gravitational
        lensing as the photons couple to Weyl tensor.

\end{abstract}

\pacs{ 04.70.Dy, 95.30.Sf, 97.60.Lf } \maketitle
\newpage
\section{Introduction}

Gravitational lensing is caused by the deflection of light ray as photons
pass close to a compact and massive body \cite{Einstein,schneider,Darwin}. Since
the images  in the gravitational lensing
carry the information about the source stars and
gravitational lens itself, gravitational lensing is a useful tool
to detect the compact astrophysical objects and to explore
the distribution of dark matter in galaxies
 \cite{Einstein,schneider,Darwin,Vir,Vir1,Vir11,Fritt,Bozza1,Eirc1,whisk,Bozza2,Bozza3,Gyulchev,sbnonk,
Bhad1,TSa1,AnAv,gr1,Kraniotis,schen,JH}.
In general, gravitational lensing depends both on the structure of
background spacetime and the dynamical
properties of the photon itself. Comparing with those in usual
 Maxwell electrodynamics,
 Eiroa \cite{Eirc2} found that photons in the non-linear
 Born-Infeld electrodynamics follow null geodesics of an
 effective metric depending
on the Born-Infeld coupling rather than of the original metric
of background spacetime, which means that the dynamics photons obeyed modifies
their propagation and changes the corresponding properties of gravitational
lensing.

The couplings between Maxwell tensor and curvature tensor also
affect the propagation of photon in background spacetime
since they change the usual Maxwell equation. Drummond \textit{et
al.}\cite{Drummond} found that these couplings could be
emerged naturally in quantum electrodynamics with the photon
effective action triggered by one-loop vacuum polarization on a
curved background spacetime. And then, he also studied the deflection angle of
the photons coupled to Riemann tensor
in the weak field limit \cite{Drummond}, which tells us
that the deflection angle of the light ray depend else on the polarization
direction of photon. This particular phenomenon stimulates a lot of interest to
probe further how these interactions between  photon and spacetime curvature tensor
affect the dynamical behavior of photon in various background spacetimes
\cite{Daniels,Caip,Cho1,Lorenci}.  In Drummond's original model, the coupling
between electromagnetic field and curvature tensor is just a quantum phenomenon,
which implies that all of the coupling constants are so tiny that
their values should be of the order of the square of the Compton wave
length of the electron
$\lambda_e$. However, the models with arbitrary coupling
constant have been investigated
widely in refs. \cite{Turner,Ni,Solanki,Dereli1,Balakin,Hehl,Bamba}
for some physical motivation. For example, Drummond's model \cite{Drummond}
 with the arbitrary coupling constant can be used to
 explain the large
scale magnetic fields observed in clusters of galaxies and the
power-law inflation in the early Universe through the special electromagnetic
fluctuations. Through adding some special coupled terms, Ni
\cite{Ni} proposed a classical generalized electromagnetic model with
the such kinds of couplings. Basing on a consideration that in the
strong gravity region closed to the classical supermassive compact
objects at the center of galaxies the coupling
between electromagnetic field and curvature tensor should be appeared
reasonably, Ni's model has been studied widely in
astrophysics \cite{Solanki,Dereli1} and black hole physics
\cite{Balakin,Hehl}.  These investigation indicate that these coupling terms
modify the equations of motion both for electromagnetic and
gravitational fields, which could lead to time delays in the arrival
of gravitational and electromagnetic waves.

Weyl tensor is one of important tensors in general relativity since it
describes a type of gravitational distortion in the spacetime.
The electrodynamics coupling with Weyl tensor
have been investigated extensively in the
literature \cite{Weyl1,Wu2011,Ma2011,Momeni,Roychowdhury,zhao2013}.
It is shown that in this  electrodynamics with Weyl correction the universal
relation with the $U(1)$ central charge is changed for the holographic conductivity
in the background of anti-de Sitter spacetime\cite{Weyl1}. Similarly,  the critical
 temperature and the order
of the phase transition are modified in the formation of the
holographic superconductor because of the presence
of such coupling terms\cite{Wu2011,Ma2011,Momeni,Roychowdhury,zhao2013}.
Moreover, we find that with these couplings the dynamical evolution
and Hawking radiation of electromagnetic field in the black hole
spacetime depend on the coupling parameter and the parity of
the field \cite{sb2013}. We also studied the effects of Weyl coupling on
the strong gravitational lensing in the background of a Schwarzschild
black hole spacetime and found that the relativistic images depend on the coupling
and the polarization direction of photon \cite{sb2015}. The time delay between
these relativistic images due to Weyl coupling is also studied in \cite{Yi101}.
Furthermore,  we studied that Weyl coupling results in the appearance of
double shadows for a single phantom black hole \cite{HS2016}. However, the above
investigation about Weyl electrodynamics are limited to the non-rotating cases.
Considering that all compact objects  are
rotating in the universe, it would be natural to think what
effects of Weyl coupling on
the strong gravitational lensing for a rotating black hole. In this paper, we will
focus on Kerr black hole and study the corresponding strong gravitational lensing
as photons couple to Weyl tensor.

The plan of our paper is organized as follows: In Sec.II,  we derive
equations of motion for the photons coupled to Weyl
tensor in the background of a Kerr
black hole from Maxwell equation with Weyl
corrections by the geometric optics approximation
\cite{Drummond,Daniels,Caip,Cho1,Lorenci}. In Sec.III, we will
study the effects of these coupling on the marginally circular photon orbit
radius, the deflection angle
and the coefficients in strong gravitational lensing in a Kerr black hole
spacetime.
In Sec.IV, we study further how Weyl
corrections and black hole rotation affect the observable including
the time delay between relativistic
images in strong gravitational lensing. We end the paper with a summary.

\section{Equation of motion for the photons coupled to Weyl tensor in a Kerr black hole spacetime}

In this section, we will make use of the geometric optics approximation
\cite{Drummond,Daniels,Caip,Cho1,Lorenci} and get
the equations of motion for the photons interacted with Weyl tensor
in a Kerr black hole spacetime.
The action for the electromagnetic field coupled to Weyl
tensor in the curved spacetime can be expressed as \cite{Weyl1}
\begin{eqnarray}
S=\int d^4x \sqrt{-g}\bigg[\frac{R}{16\pi
G}-\frac{1}{4}\bigg(F_{\mu\nu}F^{\mu\nu}-4\alpha
C^{\mu\nu\rho\sigma}F_{\mu\nu}F_{\rho\sigma}\bigg)\bigg],\label{acts}
\end{eqnarray}
which $C_{\mu\nu\rho\sigma}$ is the Weyl tensor with the form
\begin{eqnarray}
C_{\mu\nu\rho\sigma}=R_{\mu\nu\rho\sigma}-(
g_{\mu[\rho}R_{\sigma]\nu}-g_{\nu[\rho}R_{\sigma]\mu})+\frac{1}{3}R
g_{\mu[\rho}g_{\sigma]\nu}.
\end{eqnarray}
Here $g_{\mu\nu}$ is metric of the background spacetime and the brackets around indices denote the antisymmetric part. $F_{\mu\nu}$ is
the usual electromagnetic tensor and $\alpha$ is a coupling constant with
a dimension of length-squared. With the variational method,  one can find that Maxwell equation is modified as
\begin{eqnarray}
\nabla_{\mu}\bigg(F^{\mu\nu}-4\alpha
C^{\mu\nu\rho\sigma}F_{\rho\sigma}\bigg)=0,\label{WE}
\end{eqnarray}
which means that the coupling with Weyl tensor will
change propagation of electromagnetic field in background spacetime.
To get equation of motion for a coupled photon from the above
corrected Maxwell equation (\ref{WE}), we need to resort to the
geometric optics approximation. Under this approximation,  the wavelength of photon $\lambda$ is much smaller than a typical curvature scale $L$, but is
larger than the electron Compton wavelength $\lambda_e$,
 which ensures that the change of the
background gravitational and electromagnetic fields with the typical
curvature scale can be neglected for the photon
propagation \cite{Drummond,Daniels,Caip,Cho1,Lorenci}. After doing so, the electromagnetic field strength can be simplified as
\begin{eqnarray}
F_{\mu\nu}=f_{\mu\nu}e^{i\theta},\label{ef1}
\end{eqnarray}
where $f_{\mu\nu}$ and $\theta$ are, respectively, a slowly varying amplitude and a rapidly varying phase. It implies that the derivative term $f_{\mu\nu;\lambda}$ can be neglected since it is not dominated in this approximation. The wave vector is $k_{\mu}=\partial_{\mu}\theta$, which can be regarded as the usual photon momentum in quantum theories. From the Bianchi identity
\begin{eqnarray}
D_{\lambda} F_{\mu\nu}+D_{\mu} F_{\nu\lambda}+D_{\nu} F_{\lambda\mu}=0,
\end{eqnarray}
one can find that the amplitude $f_{\mu\nu}$ is constrained by
\begin{eqnarray}
k_{\lambda}f_{\mu\nu}+k_{\mu} f_{\nu\lambda}+k_{\nu} f_{\lambda\mu}=0.
\end{eqnarray}
It means that  $f_{\mu\nu}$ has a form
\begin{eqnarray}
f_{\mu\nu}=k_{\mu}a_{\nu}-k_{\nu}a_{\mu},\label{ef2}
\end{eqnarray}
where $a_{\mu}$ is the polarization vector satisfying the condition that
$k_{\mu}a^{\mu}=0$.
Combining with Eqs.(\ref{WE}), (\ref{ef1}) and (\ref{ef2}), we
can find that the equation of motion for the photon coupled to Weyl tensor has
a form
\begin{eqnarray}
k_{\mu}k^{\mu}a^{\nu}+8\alpha
C^{\mu\nu\rho\sigma}k_{\sigma}k_{\mu}a_{\rho}=0.\label{WE2}
\end{eqnarray}
It is obvious that the coupling term with Weyl tensor affects the
propagation of the coupled photon in the background spacetime.

The Kerr metric describes the gravity of a rotation black hole
in general relativity. It has a form in the standard
Boyer-Lindquist coordinates
\begin{eqnarray}
ds^2&=&-\rho^2\frac{\Delta}{\Sigma^2}dt^2+\frac{\rho^2}{\Delta}dr^2+\rho^2
d\theta^2+\frac{\Sigma^2}{\rho^2}\sin^2{\theta}(d\phi-\omega dt)^2,\label{m1}
\end{eqnarray}
with
\begin{eqnarray}
\omega&=&\frac{2aMr}{\Sigma^2},
\;\;\;\;\;\;\;\;\;\;\;\;\;\;\;\;\;\;\;\;\;\;\;\;\;\;\;\;\;\rho^2=r^2+a^2\cos^2\theta,
\nonumber\\
\Delta&=&r^2-2Mr+a^2,\;\;\;\;\;\;\;\;\;\;\;\;\;\;\;\;\Sigma^2=(r^2+a^2)^2-a^2\sin^2\theta \Delta.
\end{eqnarray}
Here the parameters $M$ and $a$ denote the mass and the angular momentum per unit mass of the black hole, respectively.
In order to introduce a local set of orthonormal frames in Kerr black hole spacetime, one can use the vierbein fields defined by
\begin{eqnarray}
g_{\mu\nu}=\eta_{ab}e^a_{\mu}e^b_{\nu},
\end{eqnarray}
where $\eta_{ab}$ is the Minkowski metric and the vierbeins
\begin{eqnarray}
e^a_{\mu}=\left(\begin{array}{cccc}
\rho\frac{\sqrt{\Delta}}{\Sigma}&0&0&-\frac{\omega\Sigma}{\rho}\sin\theta\\
0&\frac{\rho}{\sqrt{\Delta}}&0&0\\
0&0&\rho&0\\
&0&0&\frac{\Sigma}{\rho}\sin\theta
\end{array}\right)
\end{eqnarray}
with the inverse
\begin{eqnarray}
e_a^{\mu}=\left(\begin{array}{cccc}
\frac{\Sigma}{\rho\sqrt{\Delta}}&0&0&0\\
0&\frac{\sqrt{\Delta}}{\rho}&0&0\\
0&0&\frac{1}{\rho}&0\\
\frac{\omega\Sigma}{\rho\sqrt{\Delta}}&0&0&\frac{\rho}{\Sigma\sin\theta}
\end{array}\right).
\end{eqnarray}
Making use of the notation for the antisymmetric combination of
vierbeins \cite{Drummond,Daniels}
\begin{eqnarray}
U^{ab}_{\mu\nu}=e^a_{\mu}e^b_{\nu}-e^a_{\nu}e^b_{\mu},
\end{eqnarray}
one can find that the complete Weyl
tensor in Kerr black hole spacetime can be expressed as in a simple form \cite{Daniels}
\begin{eqnarray}
C_{\mu\nu\rho\sigma}&=&2\mathcal{A}\bigg(U^{01}_{\mu\nu}U^{01}_{\rho\sigma}
-U^{23}_{\mu\nu}U^{23}_{\rho\sigma}-U^{03}_{\mu\nu}U^{03}_{\rho\sigma}
+U^{12}_{\mu\nu}U^{12}_{\rho\sigma}\bigg)+2\mathcal{B}\bigg(
U^{02}_{\mu\nu}U^{02}_{\rho\sigma}-U^{13}_{\mu\nu}U^{13}_{\rho\sigma}-U^{03}_{\mu\nu}U^{03}_{\rho\sigma}
+U^{12}_{\mu\nu}U^{12}_{\rho\sigma}\bigg)\nonumber\\
&+&\mathcal{C}\bigg(U^{01}_{\mu\nu}U^{23}_{\rho\sigma}+
U^{23}_{\mu\nu}U^{01}_{\rho\sigma}-U^{03}_{\mu\nu}U^{12}_{\rho\sigma}
-U^{12}_{\mu\nu}U^{03}_{\rho\sigma}\bigg)+\mathcal{D}\bigg(
-U^{02}_{\mu\nu}U^{13}_{\rho\sigma}-
U^{13}_{\mu\nu}U^{02}_{\rho\sigma}-U^{03}_{\mu\nu}U^{12}_{\rho\sigma}
-U^{12}_{\mu\nu}U^{03}_{\rho\sigma}\bigg)\nonumber\\
&+&\mathcal{E}\bigg(
U^{01}_{\mu\nu}U^{02}_{\rho\sigma}+
U^{02}_{\mu\nu}U^{01}_{\rho\sigma}+U^{13}_{\mu\nu}U^{23}_{\rho\sigma}
+U^{23}_{\mu\nu}U^{13}_{\rho\sigma}\bigg)+\mathcal{F}\bigg(
U^{01}_{\mu\nu}U^{13}_{\rho\sigma}+
U^{13}_{\mu\nu}U^{01}_{\rho\sigma}-U^{02}_{\mu\nu}U^{23}_{\rho\sigma}
-U^{23}_{\mu\nu}U^{02}_{\rho\sigma}\bigg),
\end{eqnarray}
with
\begin{eqnarray}
\mathcal{A}&=&\frac{Mr}{\rho^6\Sigma^2}(r^2-3a^2\cos^2\theta)[
2(r^2+a^2)^2+a^2\Delta\sin^2\theta],\nonumber\\
\mathcal{B}&=&-\frac{Mr}{\rho^6\Sigma^2}(r^2-3a^2\cos^2\theta)[
(r^2+a^2)^2+2a^2\Delta\sin^2\theta],
\nonumber\\
\mathcal{C}&=&-\frac{aM\cos\theta}{\rho^6\Sigma^2}(3r^2-a^2\cos^2\theta)[
2(r^2+a^2)^2+a^2\Delta\sin^2\theta],\nonumber\\
\mathcal{D}&=&\frac{aM\cos\theta}{\rho^6\Sigma^2}(3r^2-a^2\cos^2\theta)[
(r^2+a^2)^2+2a^2\Delta\sin^2\theta],
\nonumber\\
\mathcal{E}&=&-\frac{3a^2M\sqrt{\Delta}\cos\theta}{\rho^6\Sigma^2}(3r^2-a^2\cos^2\theta)
(r^2+a^2)\sin\theta,\nonumber\\
\mathcal{F}&=&\frac{3aMr\sqrt{\Delta}}{\rho^6\Sigma^2}(3r^2-a^2\cos^2\theta)
(r^2+a^2)\sin\theta.
\end{eqnarray}
Comparing with that in Schwarzschild black hole spacetime,
 we find that the form of Weyl tensor is more complicated,
 which means the motion for the coupled photon is more
 complex in the rotating black hole spacetime.
Similarly, in order to obtain the equation of motion
for the coupled photon propagation in Kerr black hole spacetime,
 we can introduce three linear combinations of momentum components
 as in \cite{Drummond,Daniels}
\begin{eqnarray}
l_{\nu}=k^{\mu}U^{01}_{\mu\nu},\;\;\;\;\;\;\;\;\;\;
n_{\nu}=k^{\mu}U^{02}_{\mu\nu},\;\;\;\;\;\;\;\;\;\;
r_{\nu}=k^{\mu}U^{03}_{\mu\nu},\label{pvector1}
\end{eqnarray}
together with the dependent combinations
\begin{eqnarray}
&&p_{\nu}=k^{\mu}U^{12}_{\mu\nu}=\frac{1}{e_t^0k^t}\bigg[e_r^1k^rn_{\nu}
-e_{\theta}^2k^{\theta}l_{\nu}\bigg],\nonumber\\
&&m_{\nu}=k^{\mu}U^{23}_{\mu\nu}=\frac{1}{e_t^0k^t}
\bigg[e_{\theta}^2k^{\theta}r_{\nu}-(e_{t}^3k^{t}+e_{\phi}^3k^{\phi})n_{\nu}\bigg],\nonumber\\
&&q_{\nu}=k^{\mu}U^{13}_{\mu\nu}=\frac{1}{e_t^0k^t}
\bigg[e_{r}^1k^{r}r_{\nu}-(e_{t}^3k^{t}+e_{\phi}^3k^{\phi})l_{\nu}\bigg].\label{vect33}
\end{eqnarray}
These polarisation vectors are orthogonal to the wave vector $k_{\nu}$.
Contracting the equation (\ref{WE2}) with $l_{\nu}$, $n_{\nu}$, $r_{\nu}$, respectively, and making use of the relationship (\ref{vect33}), one can find that the equation of motion of the photon coupling with Weyl tensor (\ref{WE2}) can be simplified as a set of equations for three independent polarisation components $a\cdot l$, $a\cdot n$, and $a\cdot r$,
\begin{eqnarray}
\bigg(\begin{array}{ccc}
K_{11}&K_{12}&K_{13}\\
K_{21}&K_{22}&
K_{23}\\
K_{31}&K_{32}&K_{33}
\end{array}\bigg)
\bigg(\begin{array}{c}
a \cdot l\\
a \cdot n
\\
a \cdot r
\end{array}\bigg)=0,\label{Kk1}
\end{eqnarray}
with the coefficients
\begin{eqnarray}
K_{11}&=&[-(e_{t}^0)^2 + (e_t^3)^2] k^tk^t + (e_r^1)^2 k^rk^r + (e_{\theta})^2 k^{\theta}k^{\theta} + 2e_{\phi}^3 e_t^3 k^t k^{\phi} +
  (e_{\phi}^3)^2 k^{\phi}k^{\phi}\nonumber\\&-&
16\alpha \mathcal{A}\bigg[(e_{t}^0)^2k^tk^t-(e_r^1)^2 k^rk^r+(e_{\theta})^2 k^{\theta}k^{\theta}\bigg]+16\alpha \mathcal{B}\bigg[(e_t^3k^t+e_{\phi}^3k^{\phi})^2-(e_{\theta}^2)^2 k^{\theta}k^{\theta}\bigg]\nonumber\\
&+&8\alpha e_r^1e_{\theta}^2k^rk^{\theta}\bigg[\frac{
e_t^3k^t
+e_{\phi}^3k^{\phi}}{e_t^0k^t}\bigg(\mathcal{C}+2\mathcal{D}\bigg)+\mathcal{E}\bigg]+
8\alpha\frac{(e_t^3k^t+e_{\phi}^3k^{\phi})}{ e_t^0 k^t}\bigg[\mathcal{F}\bigg(2(e_t^0)^2k^tk^t-(e_r^1)^2k^rk^r\bigg)
\bigg],
\nonumber\\
K_{12}&=&16\alpha(\mathcal{A}+2\mathcal{B})e_{r}^1k^{r}e_{\theta}^2k^{\theta}-
8\alpha\mathcal{E}\bigg[(e_t^3 k^t+e_{\phi}^3k^{\phi})^2-(e_{r}^1)^2 k^{r}k^{r}+(e_{t}^0)^2 k^{t}k^{t}\bigg]
\nonumber\\&+&\frac{8\alpha(e_{t}^3k^{r}+e_{\phi}^3k^{\phi})}
{e_{t}^0k^{t}}\bigg[
\mathcal{C}\bigg((e_t^0)^2 k^tk^t-2(e_r^1)^2 k^rk^r\bigg)-\mathcal{D}\bigg((e_t^0)^2 k^tk^t+(e_r^1)^2 k^rk^r\bigg)+\mathcal{F}e_{r}^1k^{r}e_{\theta}^2k^{\theta}\bigg],
\nonumber\\
K_{13}&=&-8\alpha(e_t^3k^t+e_{\phi}^3k^{\phi})\bigg[2(\mathcal{A}+2\mathcal{B})e_{r}^1k^{r}
-\mathcal{E}e_{\theta}^2k^{\theta}\bigg]+\frac{8\alpha}{e_t^0k^t}\bigg[
\mathcal{C}e_{\theta}^2k^{\theta}\bigg(-2(e_t^0)^2k^tk^t+(e_r^1)^2 k^rk^r\bigg)\nonumber\\&-&
\mathcal{D}e_{\theta}^2k^{\theta}\bigg((e_t^0)^2k^tk^t+(e_r^1)^2 k^rk^r\bigg)
+\mathcal{F}e_r^1k^r\bigg(-(e_t^0)^2k^tk^t+(e_r^1)^2 k^rk^r-(e_{\theta}^2)^2k^{\theta}k^{\theta}\bigg)\bigg],
\end{eqnarray}
\begin{eqnarray}
K_{21}&=& 16\alpha(2\mathcal{A}+\mathcal{B})e_r^1 e_{\theta}^2k^r k^{\theta}-8\alpha\mathcal{E}\bigg[(e_t^3k^t+e_{\phi}^3k^{\phi})^2+(e_t^0)^2 k^tk^t-(e_{\theta}^2)^2 k^{\theta}k^{\theta}\bigg]
\nonumber\\&+&8\alpha\frac{(e_t^3k^t+e_{\phi}^3k^{\phi})}{e_t^0k^t}
\bigg[\mathcal{C}\bigg((e_t^0)^2k^tk^t+(e_{\theta}^2)^2 k^{\theta}k^{\theta} \bigg)+\mathcal{D}\bigg(-(e_t^0)^2 k^tk^t+2(e_{\theta}^2)^2 k^{\theta}k^{\theta}\bigg)
-\mathcal{F}e_r^1e_{\theta}^2k^rk^{\theta}\bigg],
\nonumber\\
K_{22}&=&[-(e_{t}^0)^2 + (e_t^3)^2] k^tk^t + (e_r^1)^2 k^rk^r + (e_{\theta})^2 k^{\theta}k^{\theta} + 2e_{\phi}^3 e_t^3 k^t k^{\phi} +
  (e_{\phi}^3)^2 k^{\phi}k^{\phi}\nonumber\\&+&
16\alpha\mathcal{A}\bigg[(e_t^3k^t+e_{\phi}^3k^{\phi})^2-(e_r^1)^2 k^rk^r\bigg]-16\alpha\mathcal{B}\bigg[(e_t^0)^2k^tk^t+(e_r^1)^2 k^rk^r-
(e_{\theta}^2)^2k^{\theta}k^{\theta}\bigg]
\nonumber\\&-&
8\alpha e_r^1e_{\theta}^2k^rk^{\theta}\bigg[\frac{
e_t^3k^t+e_{\phi}^3k^{\phi}}{e_t^0k^t}\bigg(2\mathcal{C}+\mathcal{D}\bigg)
-\mathcal{E}\bigg]-8\alpha
\frac{(e_t^3k^t+e_{\phi}^3k^{\phi})}{ e_t^0 k^t}\bigg[\mathcal{F}\bigg(2(e_t^0)^2k^tk^t-(e_{\theta}^2)^2k^{\theta}k^{\theta}\bigg)
\bigg],
\nonumber\\
K_{23}&=&-8\alpha(e_t^3k^t+e_{\phi}^3k^{\phi})\bigg[2(2\mathcal{A}+\mathcal{B})
e_{\theta}^2k^{\theta}-\mathcal{E}e_{r}^1k^{r}\bigg]+\frac{8\alpha}{e_t^0k^t}\bigg[
\mathcal{C}e_r^1k^r\bigg((e_t^0)^2k^tk^t+(e_{\theta}^2)^2 k^{\theta}k^{\theta}\bigg)\nonumber\\&+&
\mathcal{D}e_r^1k^r\bigg(2(e_t^0)^2k^tk^t-(e_{\theta}^2)^2 k^{\theta}k^{\theta}\bigg)
+\mathcal{F}e_{\theta}^2k^{\theta}\bigg((e_t^0)^2k^tk^t+(e_r^1)^2 k^rk^r-(e_{\theta}^2)^2k^{\theta}k^{\theta}\bigg)\bigg],
\end{eqnarray}
\begin{eqnarray}
K_{31}&=&16\alpha (e_t^3k^t+e_{\phi}^3k^{\phi})\bigg[
\frac{e_r^1k^r}{e_t^0k^t}\bigg(\mathcal{A}(e_t^0k^t-e_{\theta}^2k^{\theta})-
\mathcal{B}(e_t^0k^t+e_{\theta}^2k^{\theta})\bigg)+
\mathcal{E}e_{\theta}^2k^{\theta}\bigg]
\nonumber\\
&+&8\alpha\frac{e_{\theta}^2k^{\theta}}{e_t^0k^t}\bigg[\mathcal{C}\bigg(
-2(e_t^0)^2k^tk^t+(e_t^3k^t+e_{\phi}^3k^{\phi})^2\bigg)+\mathcal{D}\bigg(
-(e_t^0)^2k^tk^t+2(e_t^3k^t+e_{\phi}^3k^{\phi})^2\bigg)\bigg]\nonumber\\
&-&8\alpha\frac{e_{r}^1k^{r}}{e_t^0k^t}\bigg[\mathcal{F}\bigg(
(e_t^0)^2k^tk^t+(e_t^3k^t+e_{\phi}^3k^{\phi})^2\bigg)\bigg],
\nonumber\\
K_{32}&=&16\alpha (e_t^3k^t+e_{\phi}^3k^{\phi})\bigg[
\frac{1}{e_t^0k^t}\bigg(\mathcal{A}[(e_r^1)^2k^rk^r-
e_t^0k^te_{\theta}^2k^{\theta}]+\mathcal{B}[(e_r^1)^2k^rk^r+
e_t^0k^te_{\theta}^2k^{\theta}]\bigg)+\mathcal{E}e_{r}^1k^{r}\bigg]
\nonumber\\
&+&8\alpha\frac{e_{r}^1k^{r}}{e_t^0k^t}\bigg[\mathcal{C}\bigg(
(e_t^0)^2k^tk^t-2(e_t^3k^t+e_{\phi}^3k^{\phi})^2\bigg)+\mathcal{D}\bigg(
2(e_t^0)^2k^tk^t-(e_t^3k^t+e_{\phi}^3k^{\phi})^2\bigg)\bigg]\nonumber\\
&+&8\alpha\frac{e_{\theta}^2k^{\theta}}{e_t^0k^t}\bigg[\mathcal{F}\bigg(
(e_t^0)^2k^tk^t+(e_t^3k^t+e_{\phi}^3k^{\phi})^2\bigg)\bigg],\nonumber
\end{eqnarray}
\begin{eqnarray}
K_{33}&=&[-(e_{t}^0)^2 + (e_t^3)^2] k^tk^t + (e_r^1)^2 k^rk^r + (e_{\theta})^2 k^{\theta}k^{\theta} + 2e_{\phi}^3 e_t^3 k^t k^{\phi} +
  (e_{\phi}^3)^2 k^{\phi}k^{\phi}\nonumber\\&+&16\alpha \mathcal{A}\bigg[(e_{t}^0)^2 k^tk^t+(e_{\theta})^2 k^{\theta}k^{\theta} -(e_t^3k^t+e_{\phi}^3k^{\phi})^2\bigg]+16\alpha \mathcal{B}\bigg[(e_{t}^0)^2 k^tk^t+(e_{r})^2 k^{r}k^{r}-(e_t^3k^t+e_{\phi}^3k^{\phi})^2\bigg] \nonumber\\
&-&8\alpha
\frac{e_t^3k^t+e_{\phi}^3k^{\phi}}{e_t^0 k^t}\bigg[e_r^1k^r\bigg(\mathcal{C}(e_t^0 k^t-e_{\theta}^2k^{\theta})+\mathcal{D}(e_t^0 k^t+e_{\theta}^2k^{\theta})\bigg)
-\mathcal{F}\bigg((e_r^1)^2k^rk^r-(e_{\theta})^2 k^{\theta}k^{\theta} \bigg)
\bigg]\nonumber\\&-&
16\alpha\mathcal{E}e_r^1k^re_{\theta}^2 k^{\theta}.
\end{eqnarray}
The condition of Eq.(\ref{Kk1}) with non-zero solution is
 that the determinant of the matrix $K$ composed by the above
 coefficients $K_{ij}$ is equal to zero (.i.e., $|K|=0$).
 However, comparing with the case in a Schwarzschild black hole
  spacetime, we find that in the rotating black hole spacetime
  the coefficients $K_{ij}$ becomes so complicated
   that it is difficult to find a solution satisfied $|K|=0$
   in a general case. Here, we focus on only a special case in which
the whole trajectory of the photon is limited in the equatorial
plane in a Kerr black hole spacetime. For the photon propagating
 in the equatorial plane (i.e, $k^{\theta}=0$ and $\theta=\frac{\pi}{2}$),
 we find that the coefficients $K_{12}$, $K_{21}$, $K_{23}$ and $\mathcal{C}$,
 $\mathcal{D}$, $\mathcal{E}$ disappear, and then the corresponding condition
 of Eq.(\ref{Kk1}) with non-zero solution can be reduced as
\begin{eqnarray}
|K|=\tilde{K}_{11}\tilde{K}_{22}\tilde{K}_{33}=0,
\end{eqnarray}
with
\begin{eqnarray}
\tilde{K}_{11}&=&\bigg[-(e_{t}^0)^2+(e_t^3)^2\bigg]k^tk^t+(e_r^1)^2 k^rk^r+2e_{\phi}^3 e_t^3 k^t k^{\phi}+(e_{\phi}^3)^2 k^{\phi}k^{\phi},\nonumber\\
\tilde{K}_{22}&=&-(1+16\alpha\mathcal{B})(e_{t}^0)^2k^tk^t+\bigg[1-16\alpha(\mathcal{A}+
\mathcal{B})\bigg](e_{r}^1)^2k^rk^r+(1+16\alpha\mathcal{A})(e_{t}^3k^t+e_{\phi}^3k^{\phi})^2
\nonumber\\&-&16\alpha\mathcal{F}e_{t}^0k^t(e_{t}^3k^t+e_{\phi}^3k^{\phi}),
\nonumber\\
\tilde{K}_{33}&=&\bigg[1-16\alpha(\mathcal{A}+\mathcal{B})\bigg]\bigg[-(1+16\alpha\mathcal{A})(e_{t}^0)^2k^tk^t
+(1+16\alpha\mathcal{B})(e_{t}^3k^t+e_{\phi}^3k^{\phi})^2
\nonumber\\&+&16\alpha\mathcal{F}e_{t}^0k^t(e_{t}^3k^t+e_{\phi}^3k^{\phi})\bigg]
+\bigg[1+16\alpha(\mathcal{A}+\mathcal{B})+256\alpha^2\mathcal{A}\mathcal{B} +64\alpha^2 \mathcal{F}^2)\bigg](e_{r}^1)^2k^rk^r.
\end{eqnarray}
The first root $\tilde{K}_{11}=0$ leads to
\begin{eqnarray}
g_{00}k^tk^t+g_{11}k^rk^r+2g_{03}k^{t}k^{\phi}+g_{33}k^{\phi}k^{\phi}=k^{\mu}k_{\mu}=0, \label{Kcoe1}
\end{eqnarray}
which means that the light cone is not
modified by the coupling in this case.
Actually, it corresponds to an unphysical polarisation as in refs.
\cite{Drummond,Daniels,Caip,Cho1,Lorenci} and should
 be neglected  in propagation of a coupled photon.
The second root $\tilde{K}_{22}=0$,
\begin{eqnarray}
&&-(1+16\alpha\mathcal{B})(e_{t}^0)^2k^tk^t+\bigg[1-16\alpha(\mathcal{A}+
\mathcal{B})\bigg](e_{r}^1)^2k^rk^r+(1+16\alpha\mathcal{A})(e_{t}^3k^t+e_{\phi}^3k^{\phi})^2
\nonumber\\&&-16\alpha\mathcal{F}e_{t}^0k^t(e_{t}^3k^t+e_{\phi}^3k^{\phi})=0, \label{Kcoe2}
\end{eqnarray}
which corresponds to the case the polarisation vector $a_{\mu}$ is proportional to $n_{\mu}$. However, for the photon moving in the equatorial plane ($\theta=\frac{\pi}{2}$ and $k_{\theta}=0$), we find that the wave vector becomes $k_{\mu}=(k_t,k_r,0,k_{\phi})$, and the polarisation vectors $m_{\mu}$ and $n_{\mu}$ can be simplified further as
\begin{eqnarray}
m_{\mu}=(0,0, -e_{\theta}^2(e_t^3k^t+e_{\phi}^3k^{\phi}),0),\;\;\;\;\;\;n_{\mu}=(0,0,- e_{\theta}^2e_t^0k^t,0).
\end{eqnarray}
This means that the vector $n_{\mu}$ has the
 same direction as the vector $m_{\mu}$ for
 the photon limited in the equatorial plane.
 Moreover, one can find that the vector $m_{\mu}$
 denotes the polarization orthogonal to the plane
  of motion in this case. For the consistence
  with the previous study \cite{Drummond,sb2015,Yi101}, we still call the
  photon moved along the orbit (\ref{Kcoe2}) as PPM ( which stands for  a photon with a polarization along the vector $m_{\mu}$).
The third root is $\tilde{K}_{33}=0$, i.e.,
\begin{eqnarray}
&&\bigg[1-16\alpha(\mathcal{A}+\mathcal{B})\bigg]\bigg[-(1+16\alpha\mathcal{A})(e_{t}^0)^2k^tk^t
+(1+16\alpha\mathcal{B})(e_{t}^3k^t+e_{\phi}^3k^{\phi})^2
\nonumber\\&+&16\alpha\mathcal{F}e_{t}^0k^t(e_{t}^3k^t+e_{\phi}^3k^{\phi})\bigg]
+\bigg[1+16\alpha(\mathcal{A}+\mathcal{B})+256\alpha^2\mathcal{A}\mathcal{B} +64\alpha^2 \mathcal{F}^2)\bigg](e_{r}^1)^2k^rk^r=0.\label{Kcoe3}
\end{eqnarray}
This corresponds to the case the polarisation vector
$a_{\mu}$ is proportional to $K_{33}l_{\mu}-K_{31}r_{\mu}$,
 which lies on the plane of motion since $l_{\mu}$ and $r_{\mu}$
 are reduced as $(-e_r^1e_t^0k^r, e_r^1e_t^0k^r, 0, 0)$ and
 $(-e_{\phi}^3e_t^0k^{\phi}, 0, 0,e_{\phi}^3e_t^0k^t)$
 respectively when $k_{\theta}=0$. As the rotation
 parameter $a=0$, one can find the coefficient $K_{31}=0$ and
 the polarisation vector of photon becomes along $l_{\mu}$ as
 studied in previous literatures \cite{sb2015}. Similarly,
  we here still call the photon moved along the orbit (\ref{Kcoe3})
   as PPL \cite{Drummond,sb2015,Yi101}. From Eqs.(\ref{Kcoe2}) and (\ref{Kcoe3}),
we find the effects of Weyl tensor on a coupled photon propagation in a
Kerr black hole spacetime also depend on  the polarizations of photon,
which results in a phenomenon of birefringence
 \cite{Daniels,Caip,Cho1,Lorenci} as in a Schwarzschild black hole spacetime.
When the coupling constant $\alpha=0$, the
 effects of Weyl tensor vanish and then the
 light-cone conditions (\ref{Kcoe2}) and (\ref{Kcoe3})
 recover to the usual form in a Kerr spacetime without Weyl
  corrections in which the photon propagation is independent
  of the polarization directions of photon.

\section{Effects of Weyl Corrections on strong gravitational lensing in a Kerr black hole spacetime}

Let us now study the strong gravitational lensing by a Kerr black hole
as photon couples to Weyl tensor. Actually, the light cone conditions (\ref{Kcoe2}) and (\ref{Kcoe3})
indicate that the motion of the coupled photons in the equatorial plane is non-geodesic in the Kerr metric. However, these photons in the equatorial plane can be looked as moving along the null geodesics of the effective metric $\gamma_{\mu\nu}$, i.e., $\gamma^{\mu\nu}k_{\mu}k_{\nu}=0$ \cite{Breton}. The effective metric for the coupled photon limited on the equatorial plane in a Kerr black hole spacetime can be expressed as
\begin{eqnarray}
ds^2=-A(r)dt^2+B(r)dr^2+C(r)d\phi^2-2D(r)dtd\phi.\label{l01}
\end{eqnarray}
The metric coefficients $A(r)$, $B(r)$, $C(r)$ and $D(r)$ depend on the polarization directions of the coupled photon. For the PPM,
the functions $A(r)$, $B(r)$, $C(r)$ and $D(r)$ can be expressed as
\begin{eqnarray}\label{PPMmetr}
A(r)&=&\frac{r^7(r-2M)[(r^3+2Ma^2+a^2r)^2+16\alpha Mr^3]+W_0}{
r^5(r^3+2Ma^2+a^2r)^2(r^3+8\alpha M)+W_1}\nonumber\\
B(r)&=&\frac{r^5}{(r^2-2Mr+a^2)(r^3-8\alpha M)},
\nonumber\\
C(r)&=&\frac{r^3(r^3+2Ma^2+a^2r)^2[r^4(r^3+2Ma^2+a^2r)-
8\alpha M(r^4+4a^2r^2-4a^2Mr+3a^4)]}{r^5(r^3+2Ma^2+a^2r)^2(r^3+
8\alpha M)+W_1},
\nonumber\\
D(r)&=&\frac{2aMr^3(r^3+2Ma^2+a^2r)[r^4(r^3+2Ma^2+a^2r)+W_2] }{r^5(r^3+2Ma^2+a^2r)^2(r^3+8\alpha M)+W_1}
,\label{v11}
\end{eqnarray}
with
\begin{eqnarray}
W_0&=&8\alpha M a^2r^3[r^4 (7r^2+6Mr-28M^2)+4a^2r(2r^3+7Mr^2-5M^2r-4M^3)+3a^4(r^2+6Mr+4M^2)],\nonumber\\
W_1&=&-16\alpha^2M^2[8r^8-a^2r^5(29r-122M)
-a^4r^2(127r^2-212Mr-32M^2)-45a^6r(3r-2M)-45a^8],\nonumber\\
W_2&=&-2\alpha[r^4(9r-14M)-2a^2r(9r^2-Mr-8M^2)+3a^4(3r+4M)].
\end{eqnarray}
While for PPL,  these functions become
\begin{eqnarray}\label{PPLmetr}
A(r)&=&\frac{r^{10}(r-2M)[(r^3+2Ma^2+a^2r)^2-16\alpha Mr^3]+W_3}{
r^{11}(r^3+2Ma^2+a^2r)^2(r^3+8\alpha M)+W_4+8\alpha M W_1}\nonumber\\
B(r)&=&\frac{r^{10}(r^3+2Ma^2+a^2r)^2}{(r^2-2Mr+a^2)[r^5(r^3+2Ma^2+a^2r)^2(r^3+8\alpha M)+W_1]},
\nonumber\\
C(r)&=&\frac{r^6(r^3+2Ma^2+a^2r)^2[r^4(r^3+2Ma^2+a^2r)+
8\alpha M(2r^4+5a^2r^2-2a^2Mr+3a^4)]}{(r^3-
8\alpha M)[r^5(r^3+2Ma^2+a^2r)^2(r^3+
8\alpha M)+W_1]},
\nonumber\\
D(r)&=&\frac{2aMr^6(r^3+2Ma^2+a^2r)[r^4(r^3+2Ma^2+a^2r)+W_5] }{(r^3-8\alpha M)[r^5(r^3+2Ma^2+a^2r)^2(r^3+8\alpha M)+W_1]}
,\label{v12}
\end{eqnarray}
with
\begin{eqnarray}
W_3&=&-8\alpha M a^2r^6[r^4 (5r^2+6Mr-20M^2)+a^2r(7r^3+26Mr^2-16M^2r-8M^3)+3a^4(r^2+6Mr+4M^2)],\nonumber\\
W_4&=&-48\alpha^2M^2r^3[4r^8-a^2r^5(7r-46M)
-a^4r^2(41r^2-76Mr-16M^2)-15a^6r(3r-2M)-15a^8],\nonumber\\
W_5&=&2\alpha[r^4(9r-10M)+2a^2r(9r^2+Mr-4M^2)+3a^4(3r+4M)].
\end{eqnarray}
Due to the presence of rotation parameter $a$,
the coefficients of effective metric for PPM and
PPL become more complicated. As $a$ tends to zero,
the metric coefficients $A(r)$, $B(r)$, $C(r)$ and $D(r)$
in (\ref{PPMmetr}) and (\ref{PPLmetr}) can reduced,
respectively, to those of PPM and PPL in  a
Schwarzschild black hole spacetime \cite{sb2015}.
As $\alpha$ vanishes, all of the metric coefficients
$A(r)$, $B(r)$, $C(r)$ and $D(r)$ for PPM and PPL recover
to those in the Kerr case without Weyl corrections in
which the propagation behaviors of PPM are the same as those of PPL.

The null geodesics for the effective metrics  (\ref{PPMmetr}) and (\ref{PPLmetr}) can be written in a unified form
\begin{eqnarray}
\bigg(\frac{dt}{d\lambda}\bigg)_i&=&\frac{C_i(x)-J_iD_i(x)}{D_i(x)^2+A_i(x)C_i(x)},\label{u3}\\
\bigg(\frac{d\phi}{d\lambda}\bigg)_i&=&\frac{D_i(x)+J_iA_i(x)}{D_i(x)^2+A_i(x)C_i(x)},\label{u4}\\
\bigg(\frac{dx}{d\lambda}\bigg)^2_i&=&\frac{C_i(x)-2J_iD_i(x)-J^2_iA_i(x)}{B_i(x)
[D_i(x)^2+A_i(x)C_i(x)]},
\label{cedi}
\end{eqnarray}
with $i=1,~2$ corresponding to null geodesics of the coupled photons in the effective metrics  (\ref{PPMmetr}) and (\ref{PPLmetr}) respectively.
Here $x$ is related to the radial coordinate by $x=r/2M$, and the parameters $a$ and $\alpha$ are measured in the units of $2M$.
The quantity $J_i$ is the angular momentum
of the coupled photon and $\lambda$ is an affine parameter along the null geodesics.
According to the conservation of the angular momentum along the null geodesics, we can obtain the relation
between the impact parameter $u_i(x_0)$ and the distance of the closest approach of the light ray $x_0$
\begin{eqnarray}
u_i(x_0)=J_i(x_0)=\frac{-D_i(x_0)+\sqrt{D^2_i(x_0)+A_i(x_0)C_i(x_0)}}{A_i(x_0)}.
\end{eqnarray}
As the closest distance of approach $x_0$ tends to the marginally circular orbit radius $x_{ps}$ of photon, we know that the deflect angle of the light becomes unbounded large. In a stationary axially-symmetric spacetime, the marginally circular radius of photon $x_{ps}$ is defined by the biggest real root outside the horizon of the equation
\begin{eqnarray}
A_i(x)C_i'(x)-A_i'(x)C_i(x)+2J_i[A_i'(x)D_i(x)-A_i(x)D_i'(x)]=0.\label{root}
\end{eqnarray}
 \begin{figure}[ht]
\begin{center}
\includegraphics[width=7cm]{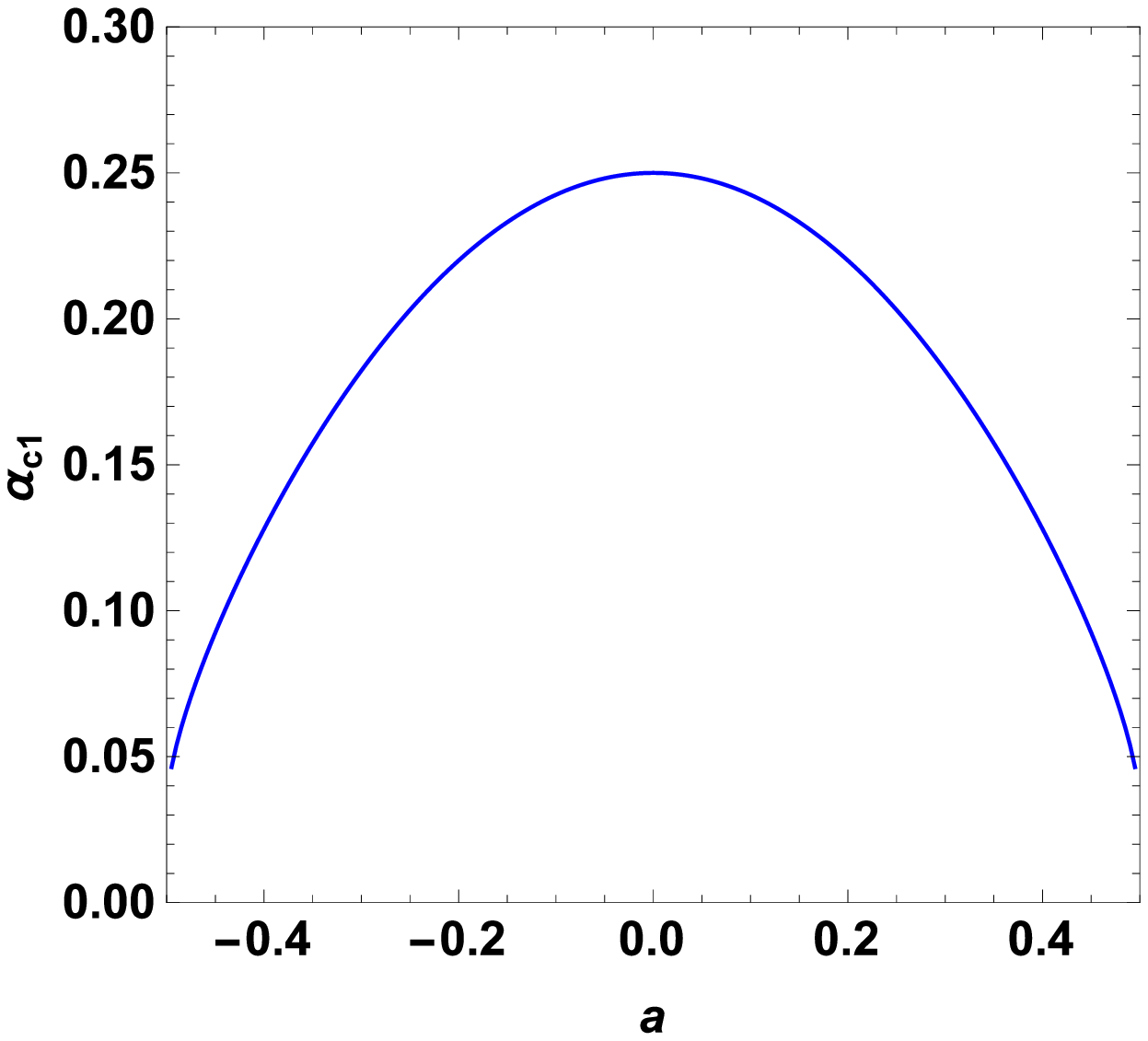}
\;\;\;\;\includegraphics[width=7cm]{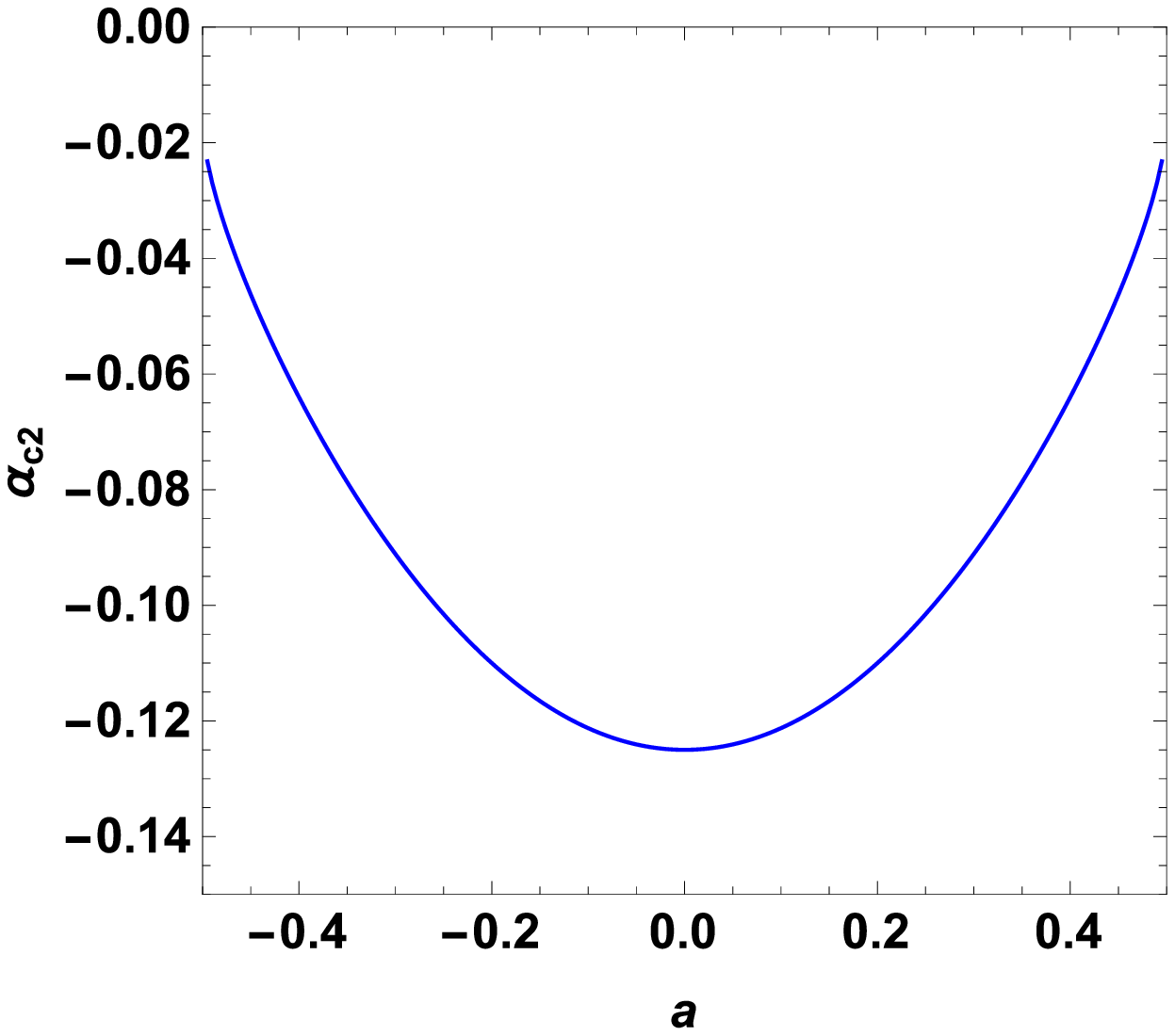}
\caption{Variety of the critical values $\alpha_{c1}$ and $\alpha_{c2}$
with the rotation parameter $a$. The marginally circular photon orbit radius $x_{ps}$ exists only in the regime $\alpha< \alpha_{c1}$ for PPM and in the regime $\alpha> \alpha_{c2}$ for PPL. Here, we
set $2M=1$.}
\end{center}
\end{figure}
Solving Eq.(\ref{root}) together with Eqs.(\ref{PPMmetr}) and (\ref{PPLmetr}), we can get the marginally circular photon orbit radius $x_{ps}$ for PPM and PPL. However, the coupling terms with Weyl tensor make the equation so complicated that it is impossible to obtain an analytical form for the marginally circular photon orbit radius and we must resort to numerical method  in this case. Our results show that the marginally circular photon orbit radius $x_{ps}$ exists only in the regime $\alpha< \alpha_{c1}$ for PPM and in the regime $\alpha> \alpha_{c2}$ for PPL. The critical value is determined by the condition that the marginally circular orbit radius for the coupled photon is overlapped with the event horizon of the black hole.
The critical values $\alpha_{c1}$ and $\alpha_{c2}$ depend on the rotation parameter $a$ and their forms can be expressed as
\begin{eqnarray}
\alpha_{c1}=-2\alpha_{c2}=\frac{1}{8}
\bigg[4M^2-3a^2+(4M^2-a^2)\sqrt{1-\frac{a^2}{M^2}}\bigg].
\end{eqnarray}
It is obvious that the critical value  $\alpha_{c1}$ decreases with $|a|$ and $\alpha_{c2}$ increases  with $|a|$, which is also shown in Fig.(1). When $a$ tends to zero, we find that $\alpha_{c1}=M^2$ and $\alpha_{c2}=-M^2/2$, which is consistent with those in Schwarzschild black hole spacetime \cite{sb2015}.
In Figs.(2) and (3), we present the dependence of the marginally circular photon orbit radius $x_{ps}$ on the coupling parameter $\alpha$ and the rotation parameter $a$ for PPM and PPL, respectively. It tells us that
that the marginally circular photon orbit radius $x_{ps}$ for PPM decreases  monotonously with the coupling parameter $\alpha$ for different $a$. In the case of PPL, we find that $x_{ps}$ for the retrograde photon increases  monotonously with $\alpha$ for all $a$. For the prograde PPL, the monotonicity of $x_{ps}$ with $\alpha$ is destroyed gradually with increasing $a$. In the case with the larger $a$, we find that $x_{ps}$ for the prograde PPL first increases with $\alpha$ and then decrease and finally increases again,  which is different from that in the Schwarzschild case \cite{sb2015}. Moreover, in the case with the larger $a$, we find from Figs.2 and 3 that there exist a distinct turning point for the change rate of $x_{ps}$ with $\alpha$. For example, as $a=0.4$, the turning point is near $\alpha\sim0.11$ for the prograde PPL and near $\alpha\sim-0.04$ for the prograde PPM. This implies that the significantly changes could be occurred  in gravitational lensing for the prograde coupled photon as the value of $\alpha$ is near this turning point. Moreover, with the increasing rotation parameter $a$, $x_{ps}$ decreases for the prograde photons with different polarization directions. However, for the retrograde photon, $x_{ps}$ increases with $a$ as the value of $\alpha$ is near the threshold value $\alpha_{c1}$ for PPM or $\alpha_{c2}$ for PPL.  Therefore, the presence of the coupling modifies the dependence of $x_{ps}$ on $a$ since in the case without this Weyl coupling a common feature of $x_{ps}$ in a rotating black hole spacetime is that it decreases monotonously with the rotation parameter $a$.
 \begin{figure}[ht]
\begin{center}
\includegraphics[width=7cm]{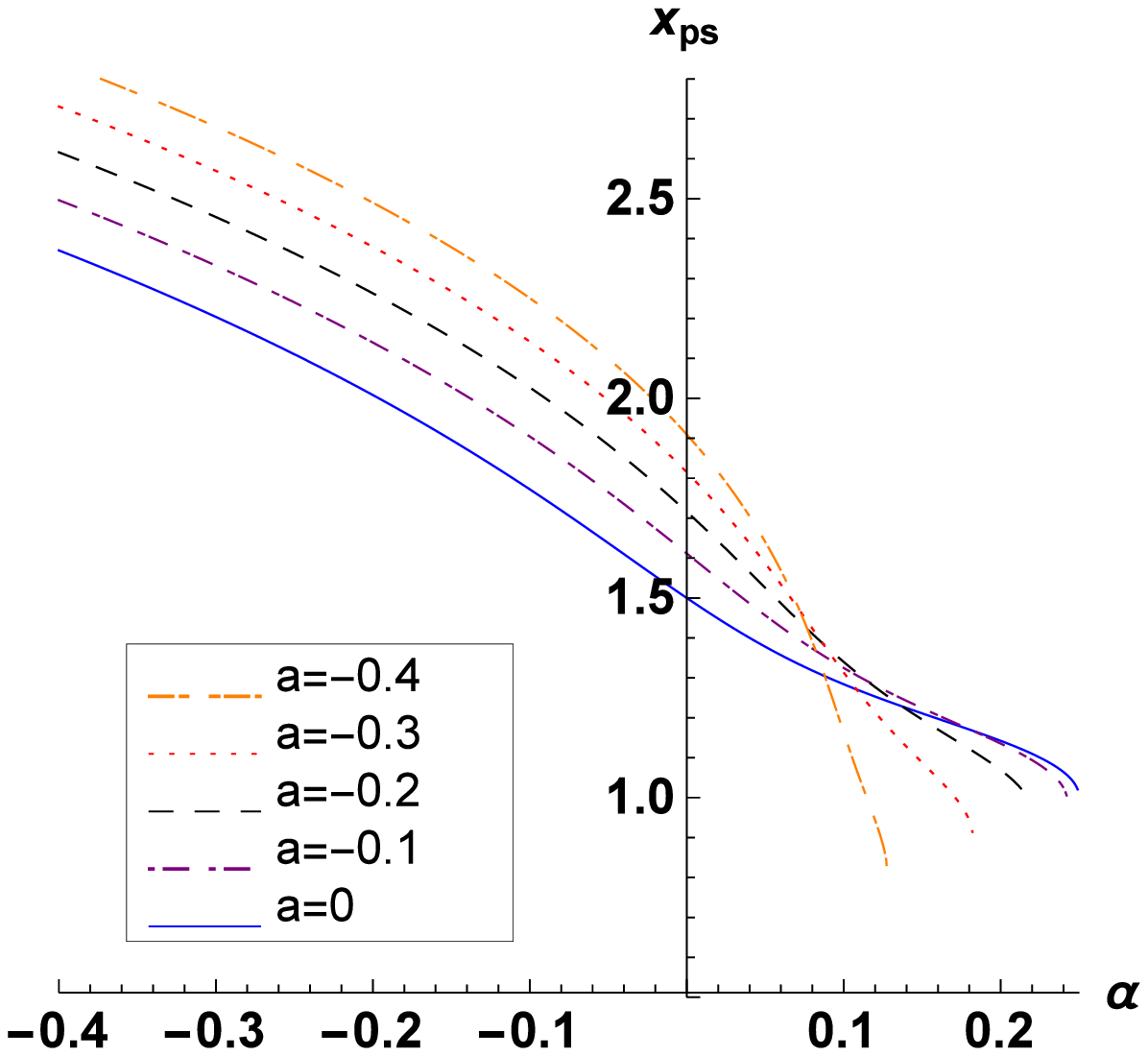}
\;\;\;\;\includegraphics[width=7cm]{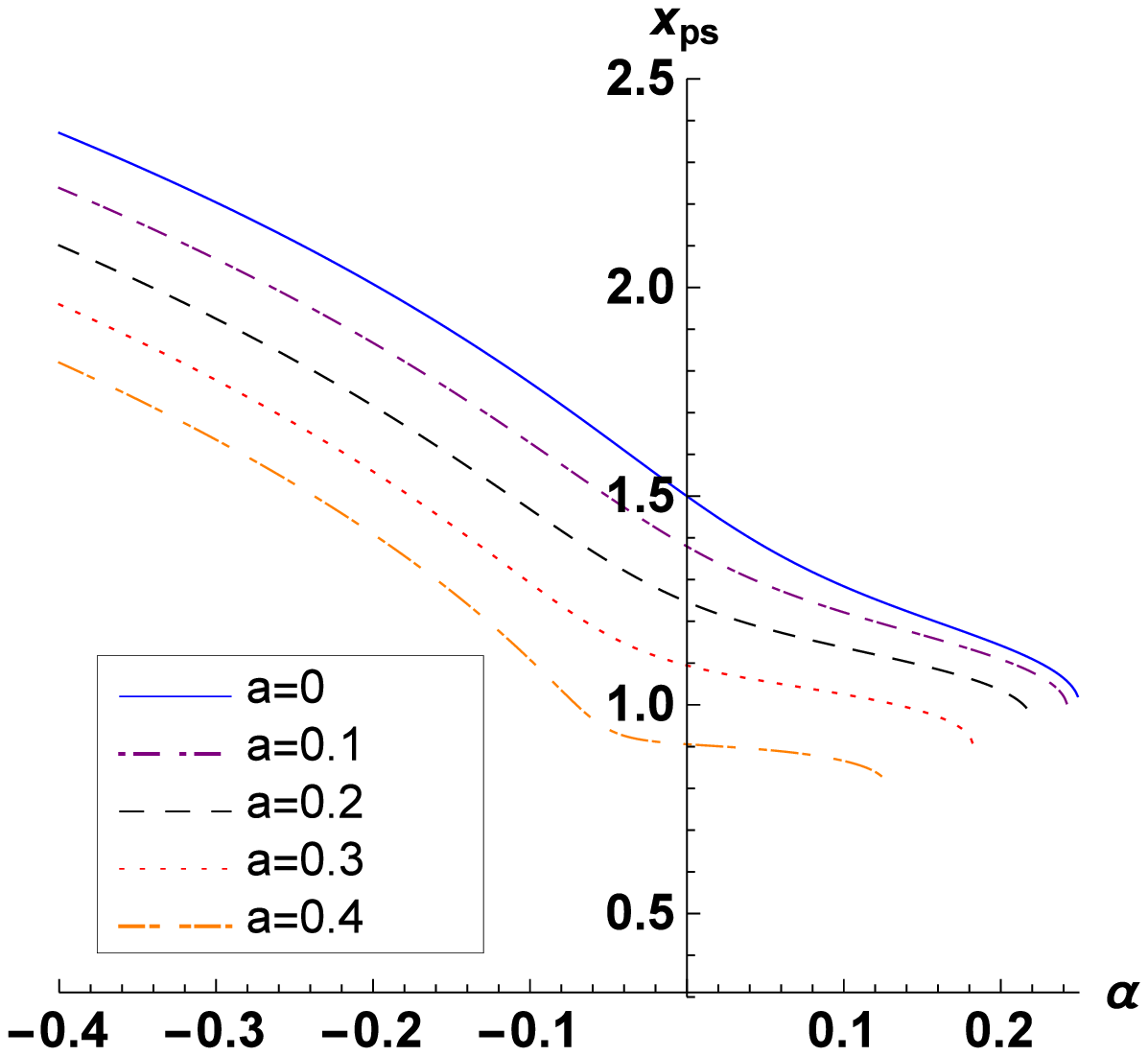}
\caption{Variety of the marginally circular orbit radius of PPM
with the coupling parameter $\alpha$ for different $a$. Here, we
set $2M=1$.}
\end{center}
\end{figure}
\begin{figure}[ht]
\begin{center}
\includegraphics[width=7cm]{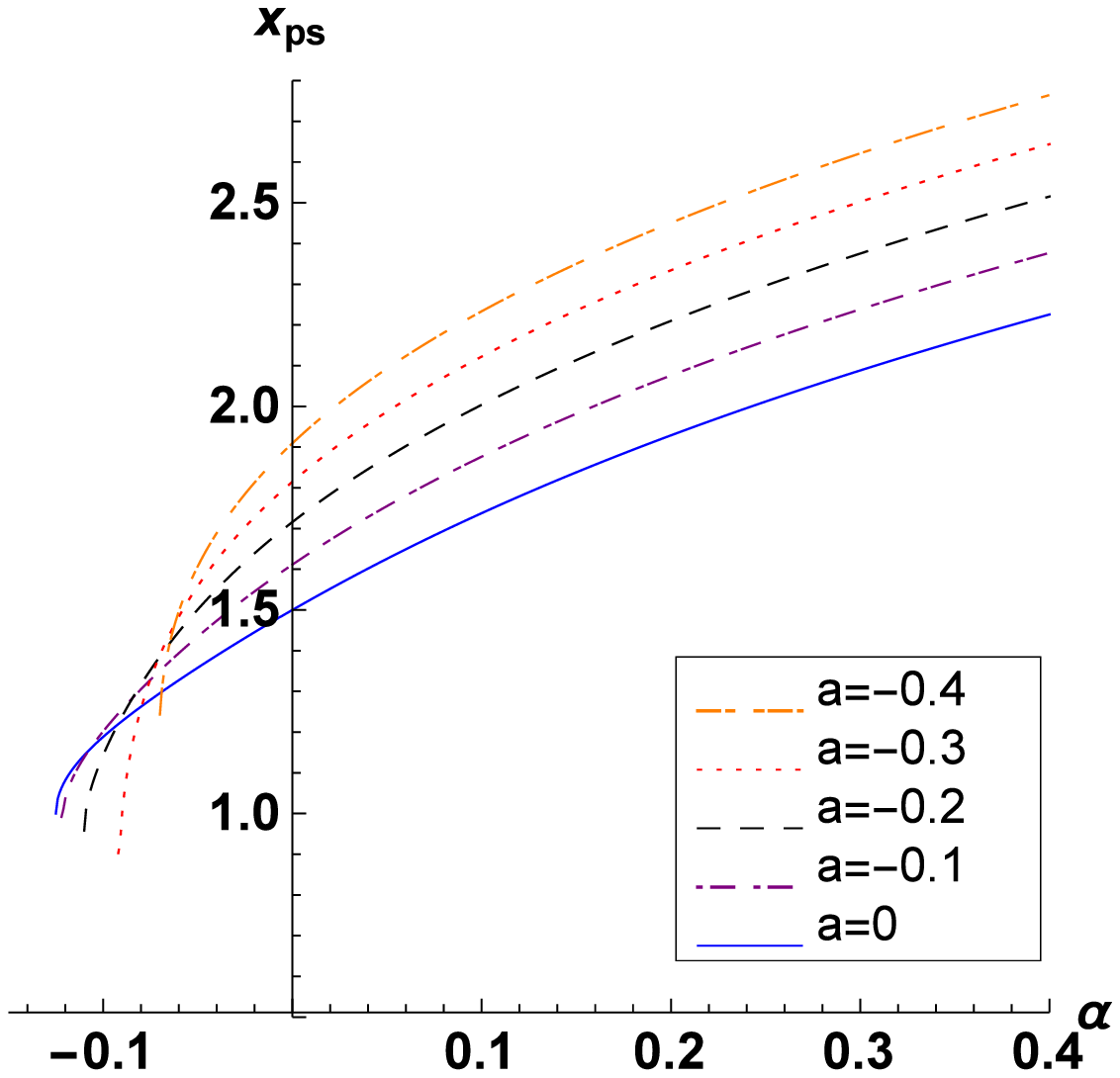}
\;\;\;\;\includegraphics[width=7cm]{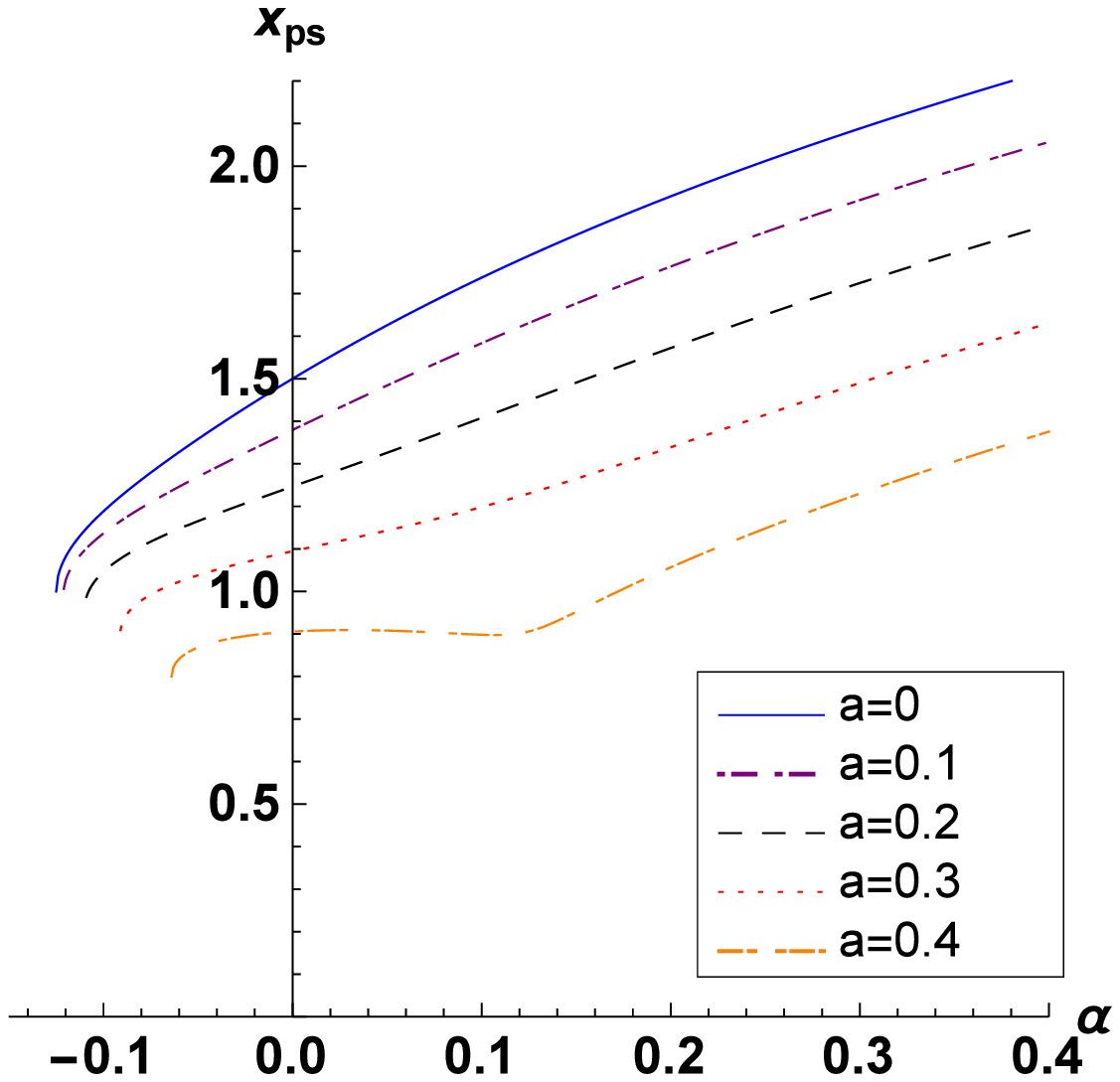}
\caption{Variety of the marginally circular orbit radius of photon PPL
with  the coupling parameter $\alpha$ for different $a$. Here, we
set $2M=1$.}
\end{center}
\end{figure}
The modification to $x_{ps}$ from the coupling implies that strong
gravitational lensing for the coupled photon in a Kerr black hole
spacetime could possess some new behaviors which is not observed
 else in the case without Weyl correction.

From the geodesic equations (\ref{cedi}) limited on the equatorial
plane, we can obtain that the deflection angle for the coupled photon
in a Kerr black hole spacetime, which has a similar form in the case
without coupling \cite{Vir1}, i.e.,
\begin{eqnarray}
\alpha_i(x_{0})=I_i(x_{0})-\pi.
\end{eqnarray}
with
\begin{eqnarray}
I_i(x_0)=2\int^{\infty}_{x_0}\frac{\sqrt{B_i(x)|A_i(x_0)|}
[D_i(x)+J_iA_i(x)]dx}{\sqrt{D_i^2(x)+A_i(x)C_i(x)}
\sqrt{A_i(x_0)C_i(x)-A_i(x)C_i(x_0)+2J_i[A_i(x)D_i(x_0)-A_i(x_0)D_i(x)]}}.
\label{int1}
\end{eqnarray}
Similarly, here $x_0$ is the closest approach distance of photon
from black hole in the propagation process. However, in this case,
the function $I_i(x_{0})$ depends on the polarization directions
 of the coupled photon since the effective metric functions
 $A_i(x)$, $B_i(x)$, $C_i(x)$ and $D_i(x)$ are
 different for the photons with different types of polarizations.
Thus, it is natural to expect that the deflection angle of
PPL has some properties differed from that of PPM.
Making use of the approximation method for the integral (\ref{int1})
proposed firstly by Bozza \cite{Bozza2}, we can study analytically
the properties of the deflection angle as photon orbit is near the
marginally circular orbit. Defining a variable
\begin{eqnarray}
z=1-\frac{x_0}{x},
\end{eqnarray}
we can rewrite the integral (\ref{int1}) as
\begin{eqnarray}
I_i(x_0)=\int^{1}_{0}R_i(z,x_0)f_i(z,x_0)dz,\label{in0}
\end{eqnarray}
with
\begin{eqnarray}
R_i(z,x_0)&=&\frac{2x^2}{x_0\sqrt{C_i(z)}}\frac{\sqrt{B_i(z)|A_i(x_0)|}
[D_i(z)+J_iA_i(z)]}{\sqrt{D_i^2(z)+A_i(z)C_i(z)}}, \label{R10}
\end{eqnarray}
\begin{eqnarray}
f_i(z,x_0)&=&\frac{1}{\sqrt{A_i(x_0)-A_i(z)\frac{C_i(x_0)}{C_i(z)}
+\frac{2J_i}{C_i(z)}[A_i(z)D_i(x_0)-A_i(x_0)D_i(z)]}}.
\end{eqnarray}
The functions $R_i(z, x_0)$ are regular for all values of $z$
and $x_0$, but the function $f_i(z, x_0)$ are divergent as $z$ tends to zero.
In order to study the dominated feature of the deflection angle
as the coupled photon is close to the marginally circular
photon orbit (.i.e.,$z\rightarrow 0$), one can
take apart the integral (\ref{in0}) into the
divergent part $I_{iD}(x_0)$ and the regular part $I_{iR}(x_0)$ for two types of
photons with different polarizations
\begin{eqnarray}
I_{iD}(x_0)&=&\int^{1}_{0}R_i(0,x_{ps})f_{i0}(z,x_0)dz, \nonumber\\
I_{iR}(x_0)&=&\int^{1}_{0}[R_i(z,x_0)f_i(z,x_0)-R_i(0,x_{ps})f_{i0}(z,x_0)]dz
\label{intbr}.
\end{eqnarray}
And then, by expanding the argument of the square root in $f_i(z,x_{0})$ to the second order in $z$, one can obtain the forms of $f_{i0}(z,x_{0})$ in Eq.(\ref{intbr})
\begin{eqnarray}
f_0(z,x_0)=\frac{1}{\sqrt{p(x_0)z+q(x_0)z^2}},\label{f0z}
\end{eqnarray}
with
\begin{eqnarray}
p_i(x_0)&=&\frac{x_0}{C_i(x_0)}\bigg\{A_i(x_0)C_i'(x_0)
-A_i'(x_0)C_i(x_0)+2J_i[A_i'(x_0)D_i(x_0)-A_i(x_0)D_i'(x_0)]\bigg\},  \nonumber\\
q_i(x_0)&=&\frac{x_0}{2C_i^2(x_0)}\bigg\{2\bigg(C_i(x_0)-x_0C_i'(x_0)\bigg)
\bigg([A_i(x_0)C_i'(x_0)-A_i'(x_0)C_i(x_0)]
+2J_i[A_i'(x_0)D_i(x_0)-A_i(x_0)D_i'(x_0)]\bigg)\nonumber\\&&+x_0C_i(x_0)
\bigg([A_i(x_0)C_i''(x_0)-A_i''(x_0)C_i(x_0)]+2J_i[A_i''(x_0)D_i(x_0)
-A_i(x_0)D_i''(x_0)]\bigg)\bigg\}.\label{al0}
\end{eqnarray}
\begin{figure}[ht]
\begin{center}
\includegraphics[width=7cm]{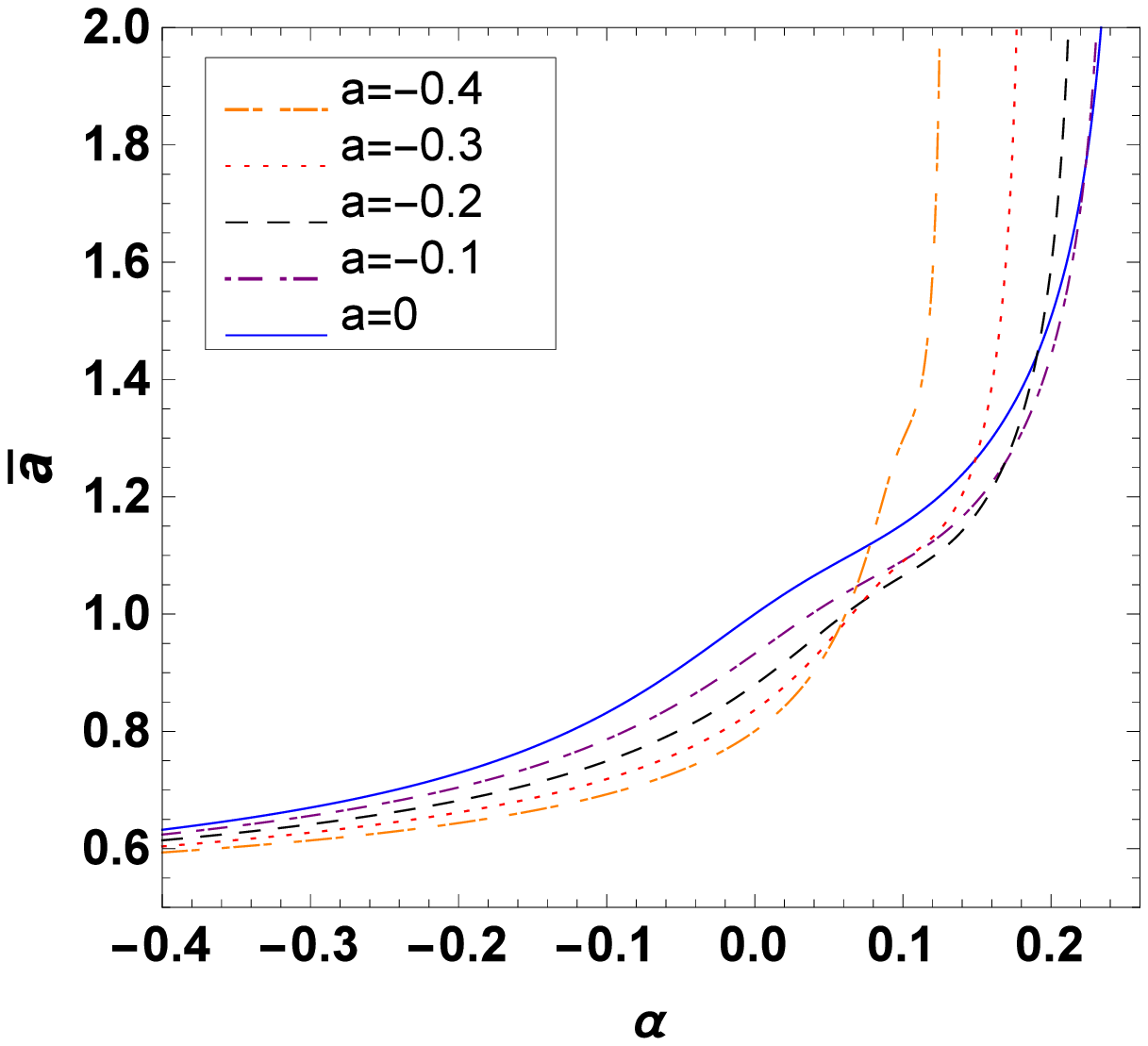}\;\;\;\;
\includegraphics[width=7cm]{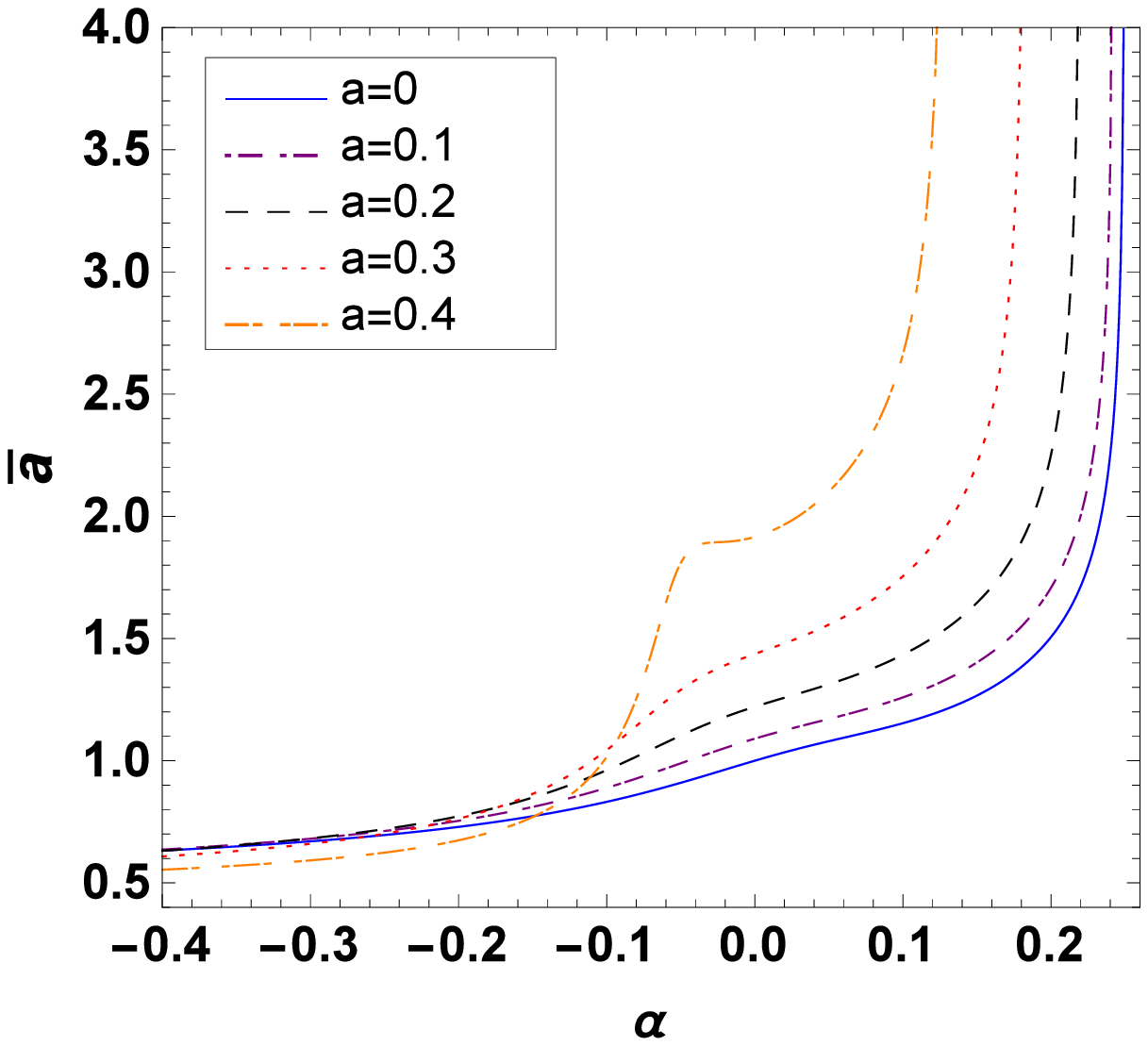}\\
\includegraphics[width=7cm]{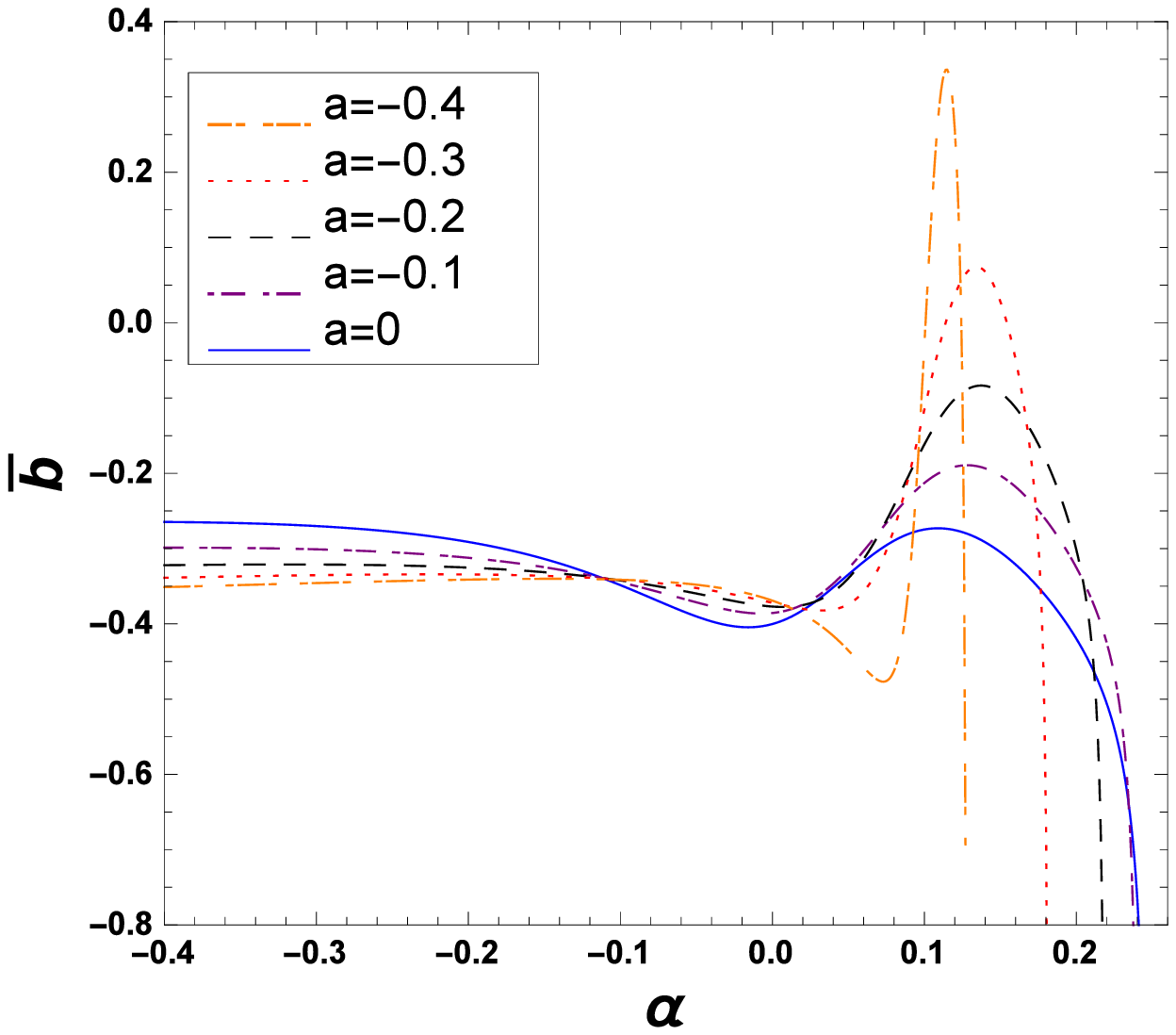}\;\;\;\;
\includegraphics[width=7cm]{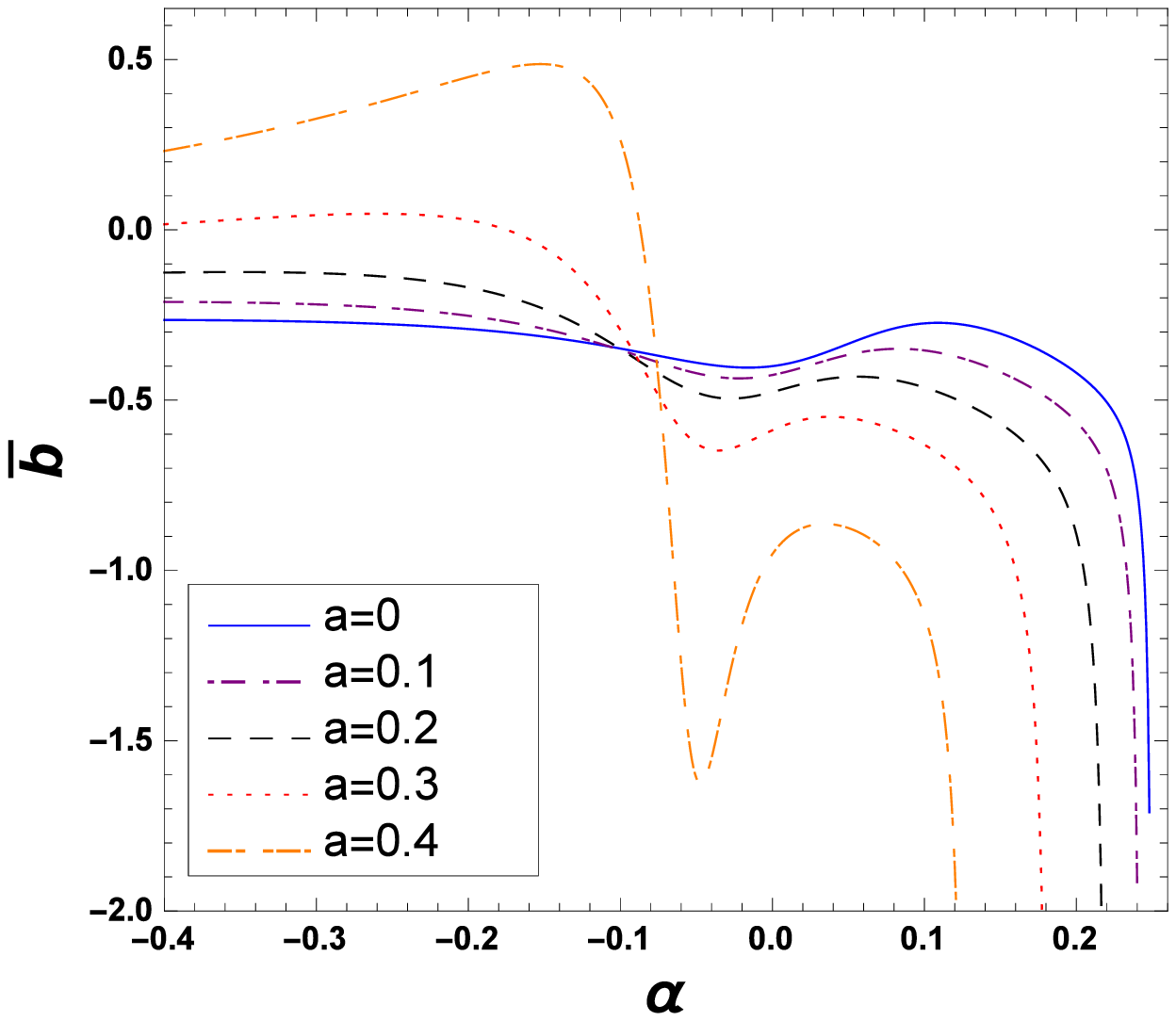}
\caption{Change of the strong deflection limit coefficients $\bar{a}$ and $\bar{b}$ for PPM with the coupling parameter $\alpha$ for different $a$ in a Kerr black hole spacetime. Here, we set $2M=1$.}
\end{center}
\end{figure}
\begin{figure}
\begin{center}
\includegraphics[width=7cm]{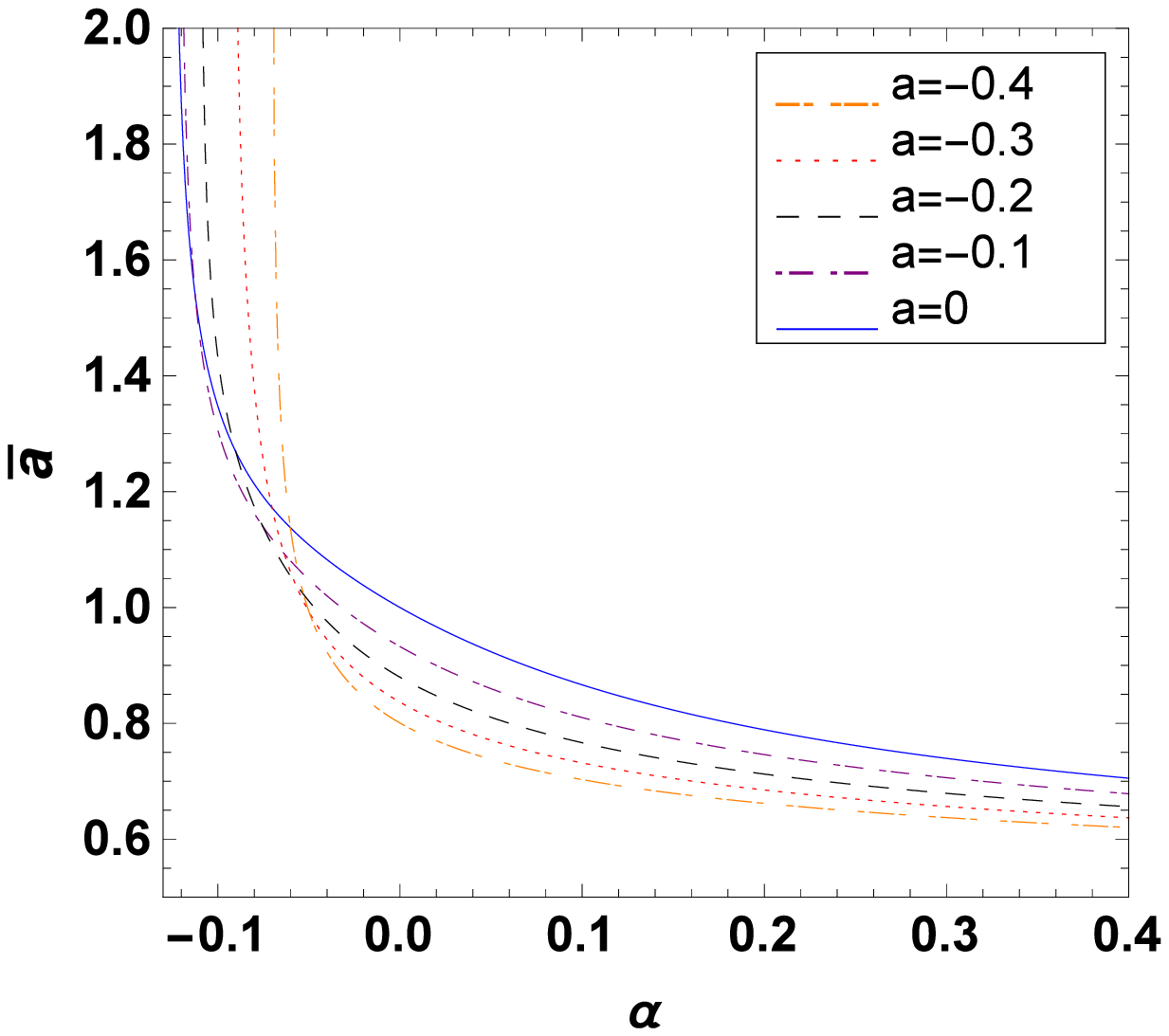}\;\;\;\;
\includegraphics[width=7cm]{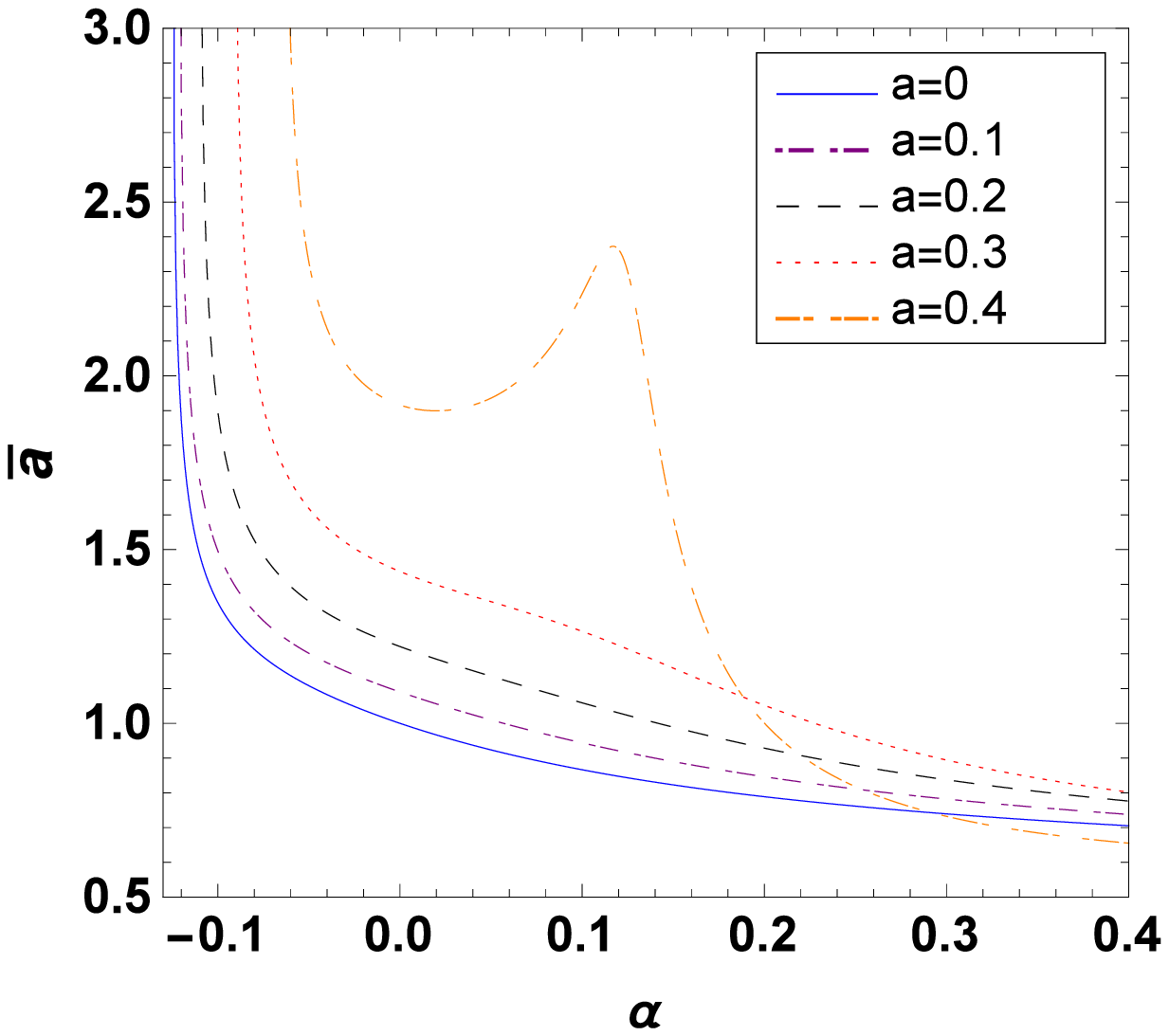}\\
\includegraphics[width=7.3cm]{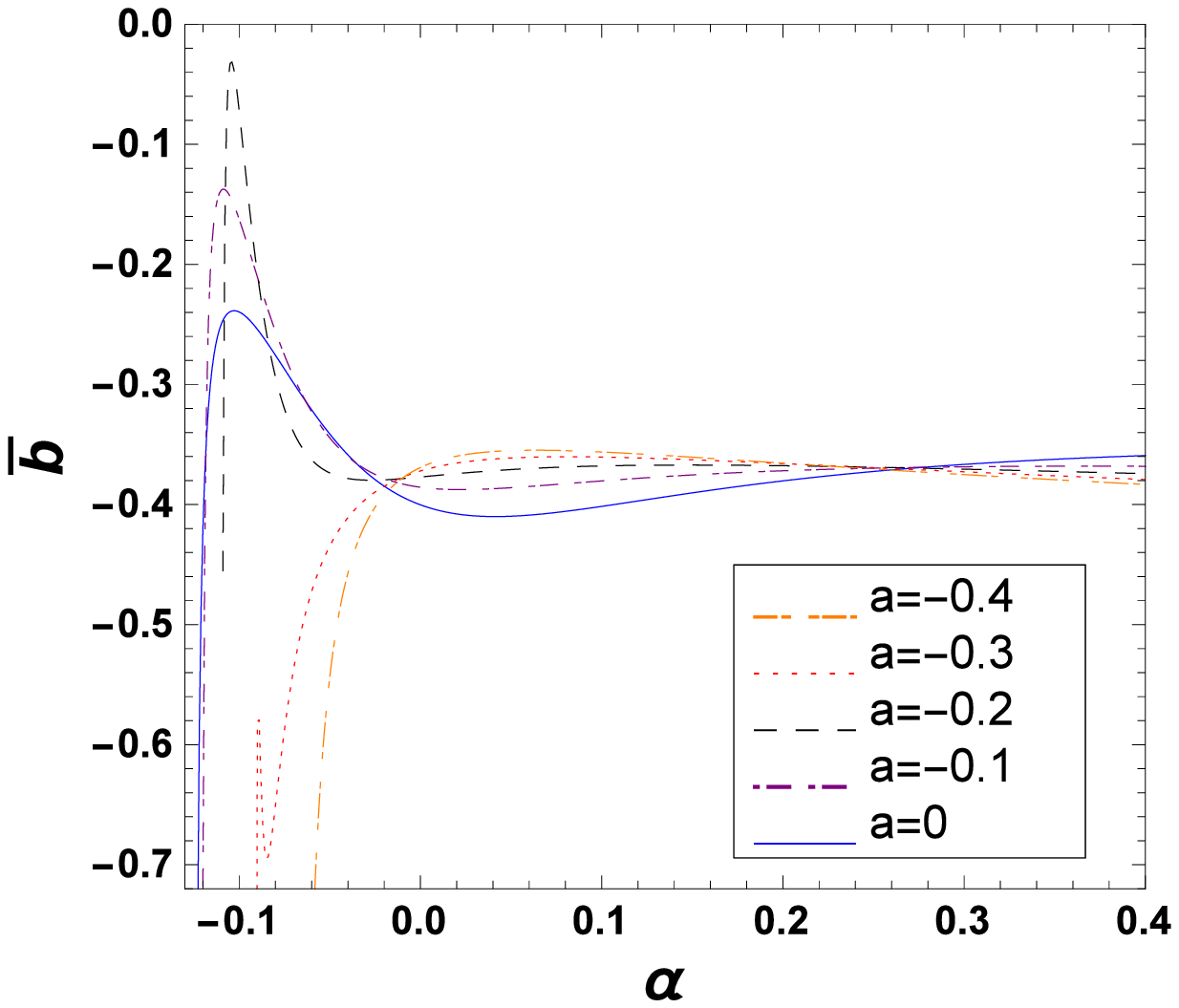}\;\;\;\;
\includegraphics[width=7cm]{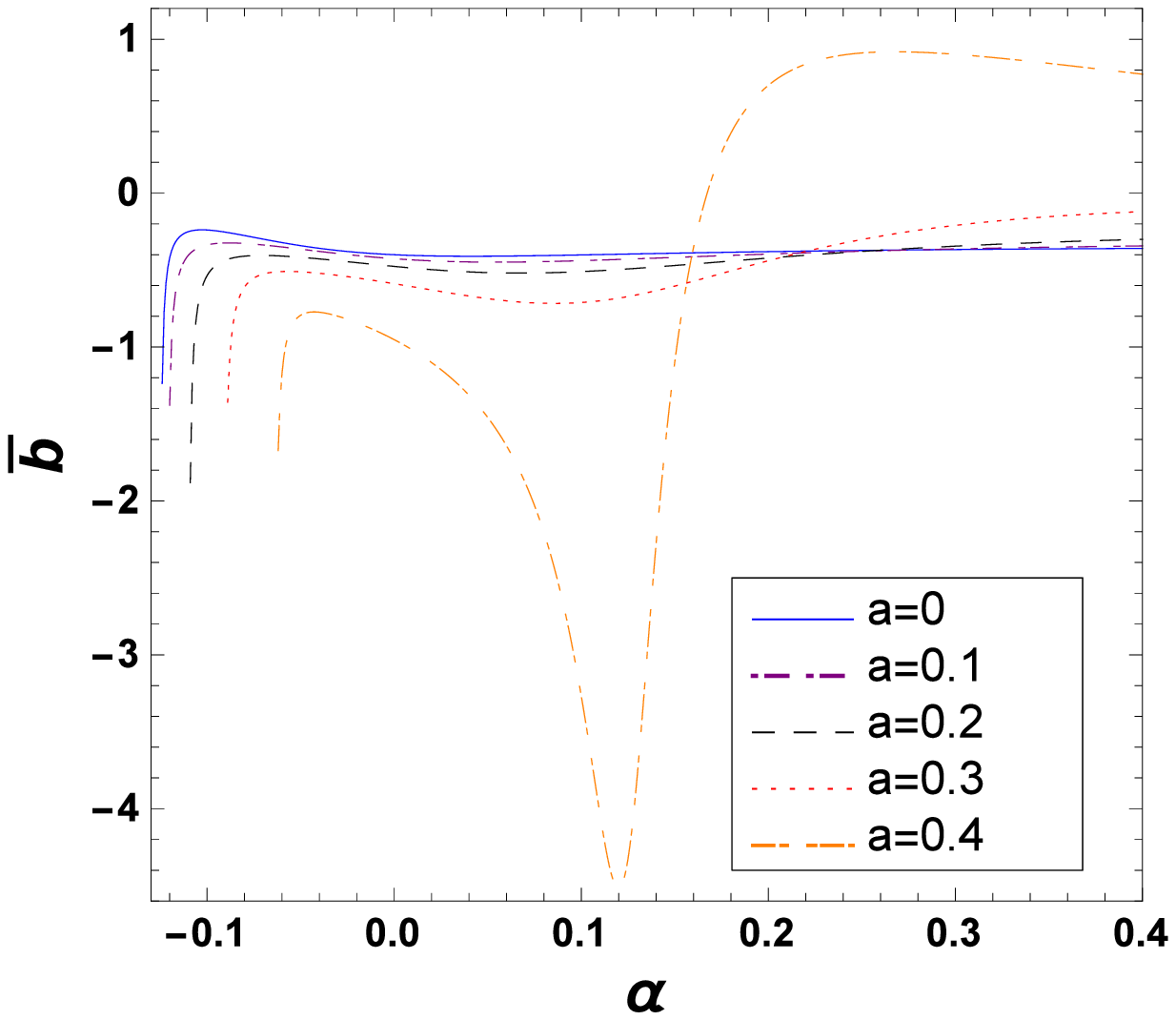}
\caption{Change of the strong deflection limit coefficients $\bar{a}$
 and $\bar{b}$ for PPL with the coupling parameter $\alpha$ for
 different $a$ in a Kerr black hole spacetime. Here, we set $2M=1$.}
\end{center}
\end{figure}
If $x_0$ tends to the radius of the marginally circular photon
orbit $x_{ps}$, from Eq.(\ref{root}), one can find that $p_i(x_{0})$ approach zero and the behaviors of $f_{i0}(z,x_{0})$ are dominated by the term $z^{-1}$ , which means that in the strong gravitational limit the deflection angle for the coupled photon diverges logarithmically \cite{Bozza2}, i.e.,
\begin{eqnarray}
\alpha_i(\theta)=-\bar{a_i}\log{\bigg(\frac{\theta
D_{OL}}{u_{ips}}-1\bigg)}+\bar{b_i}+\mathcal{O}(u-u_{ips}), \label{alf1}
\end{eqnarray}
with
\begin{eqnarray}
&\bar{a_i}&=\frac{R_i(0,x_{ps})}{\sqrt{q_i(x_{ps})}}, \nonumber\\
&\bar{b_i}&= -\pi+b_{iR}+\bar{a_i}\log{\bigg\{\frac{2q_i(x_{ps})C_i(x_{ps})}{
u_{ips}A_i(x_{ps})[D_i(x_{ps})+J_iA_i(x_{ps})]}\bigg\}}, \nonumber\\
&b_{iR}&=I_{iR}(x_{ps}), \nonumber\\
&u_{ips}&=\frac{-D_i(x_{ps})+\sqrt{A_i(x_{ps})C_i(x_{ps})+D_i^2(x_{ps})}}{
A_i(x_{ps})}.\label{coa1}
\end{eqnarray}
\begin{figure}[ht]
\begin{center}
\includegraphics[width=7cm]{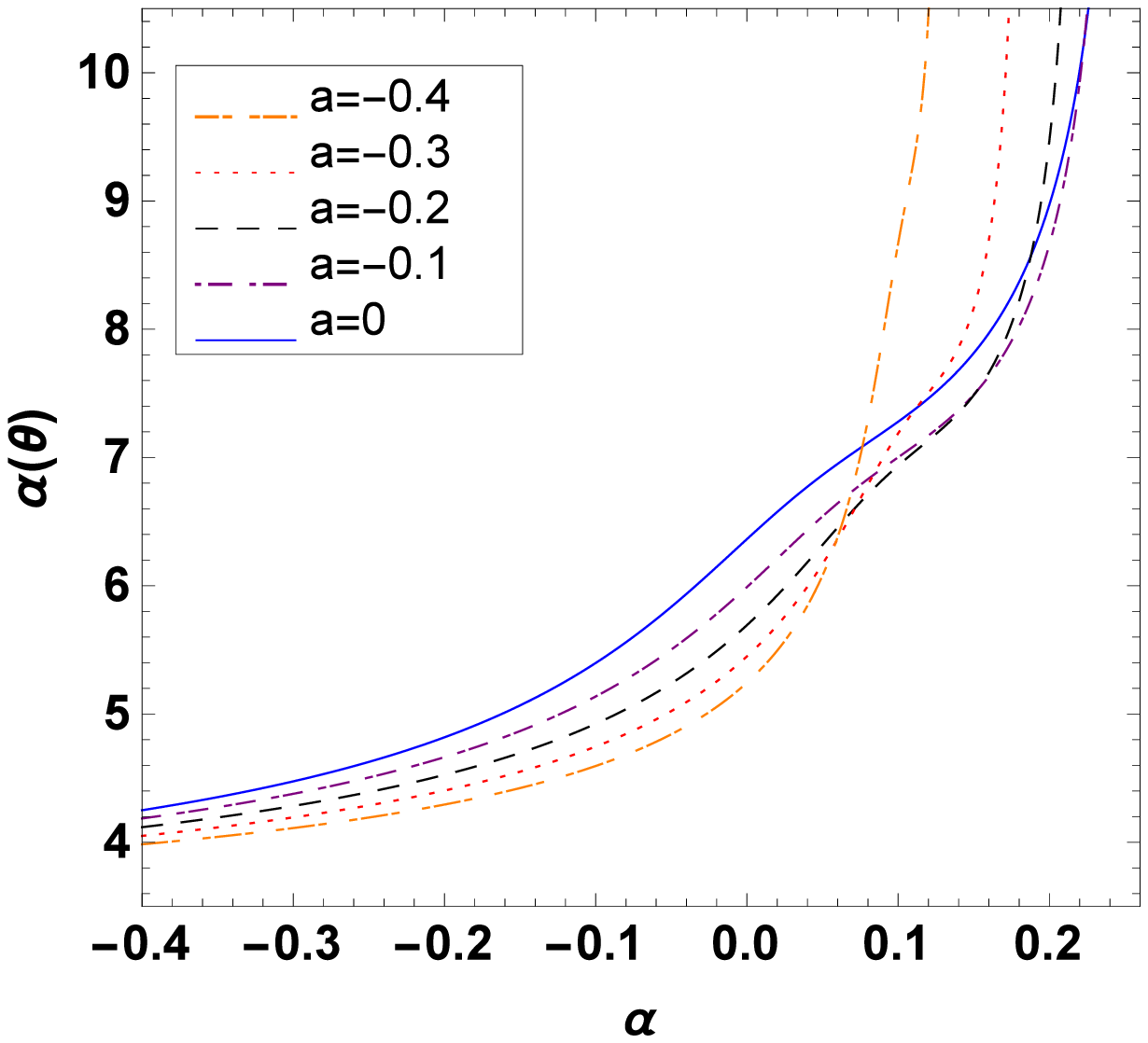}
\;\;\;\;\includegraphics[width=7cm]{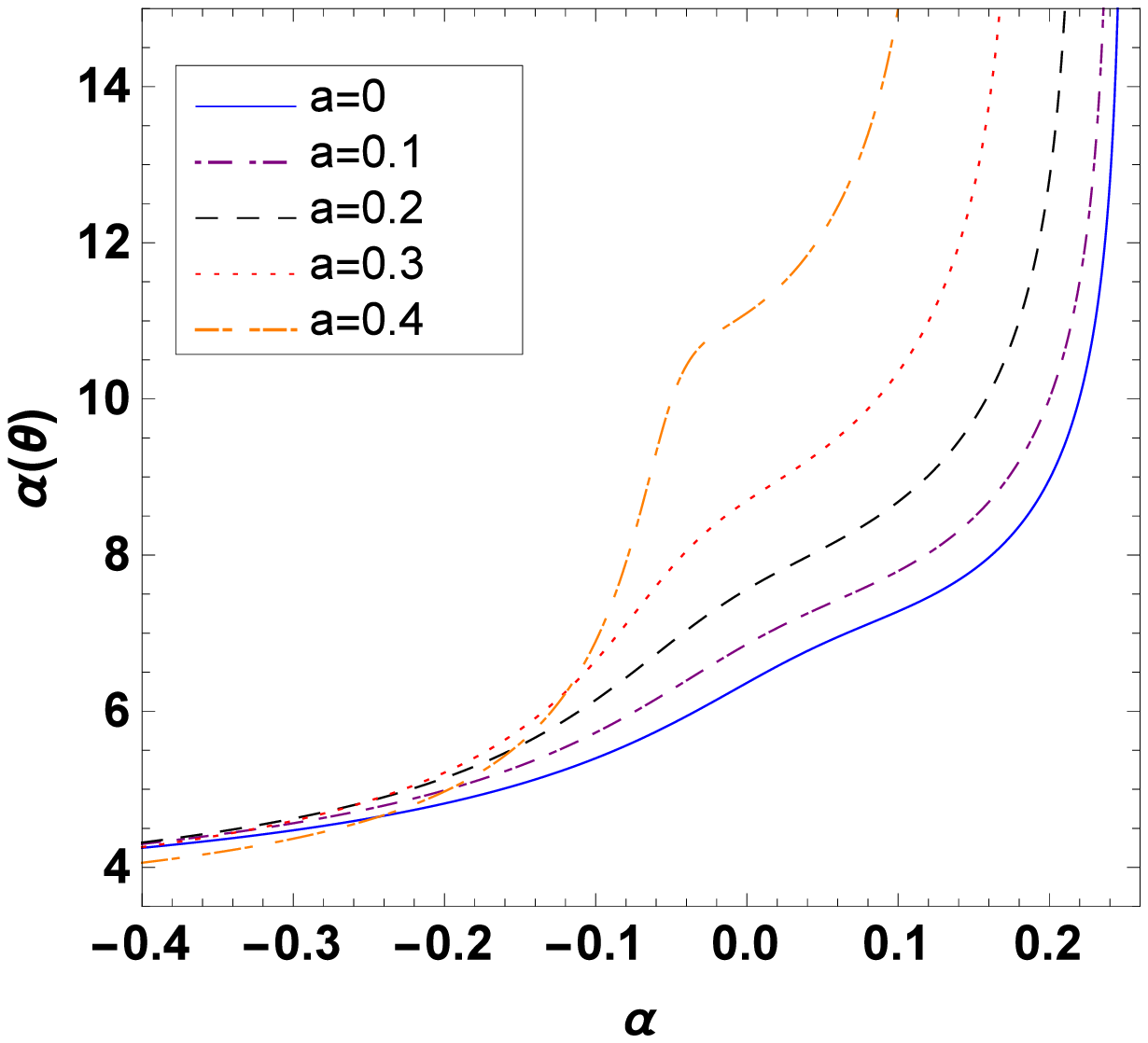}
\caption{Deflection angles for PPM evaluated at
$u=u_{ps}+0.003$ is a function of the coupling parameter
 $\alpha$ for different $a$ in a Kerr black hole spacetime. Here, we set $2M=1$.}
\end{center}
\end{figure}
\begin{figure}[ht]
\begin{center}
\includegraphics[width=7cm]{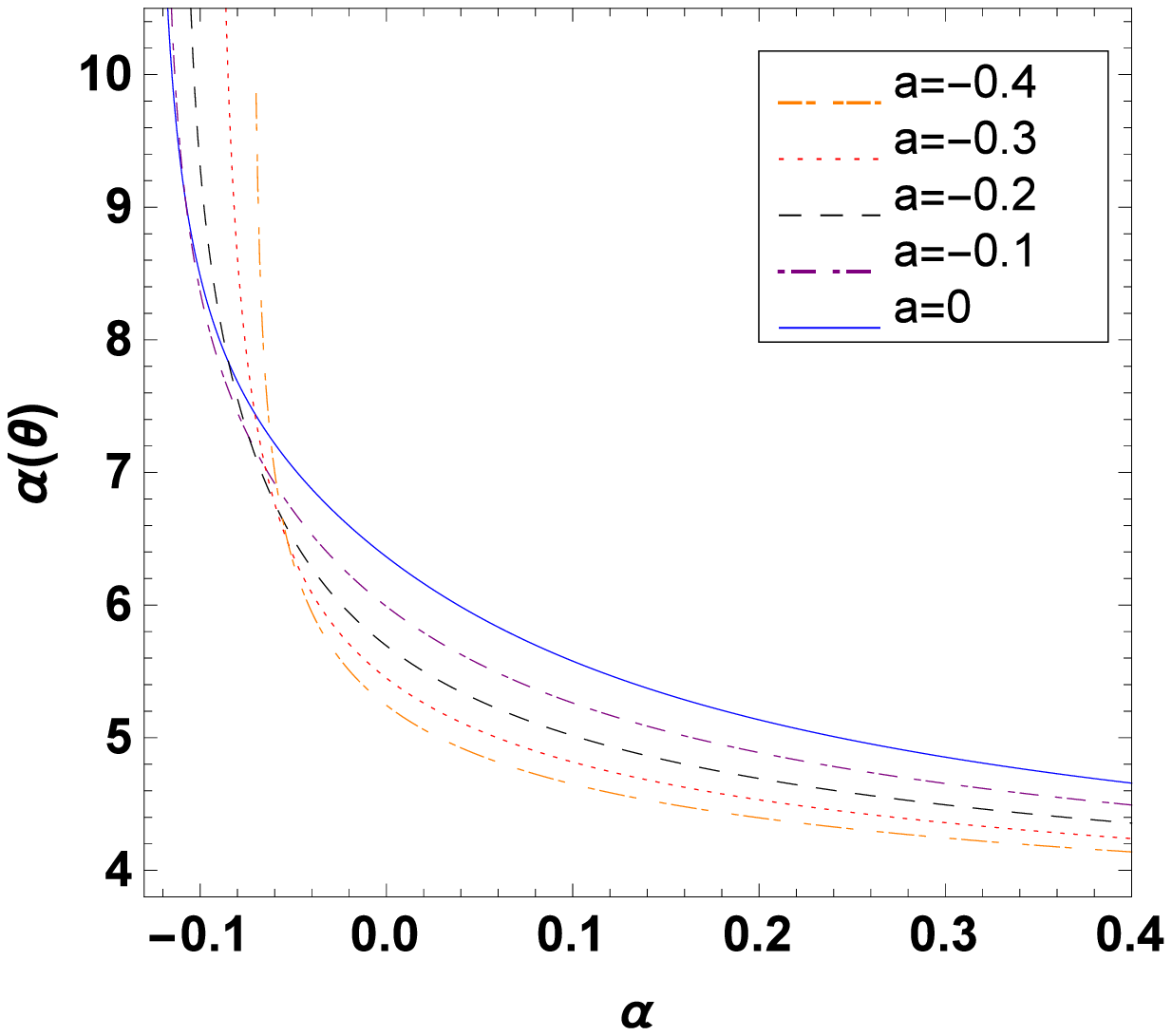}\;\;\;\;
\includegraphics[width=7cm]{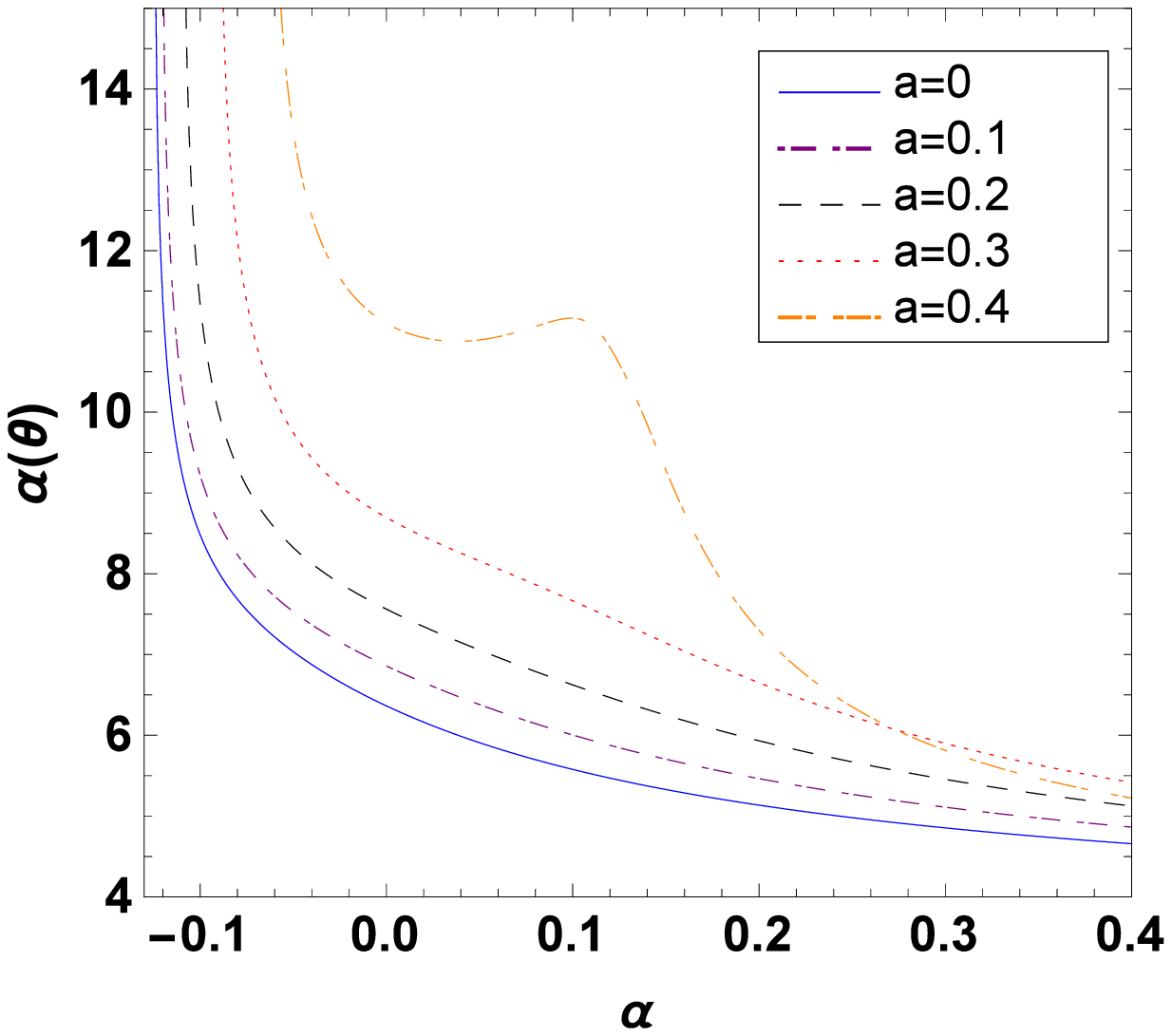}
\caption{Deflection angles for PPL evaluated at $u=u_{ps}+0.003$
is a function of the coupling parameter $\alpha$ for different $a$ in a Kerr black hole spacetime. Here, we set $2M=1$.}
\end{center}
\end{figure}
Here $D_{OL}$ denotes the distance between observer and
gravitational lens. With Eqs.(\ref{alf1}) and (\ref{coa1}),
we can study numerically the properties of strong gravitational lensing for the coupled photon in a Kerr black hole spacetime. The changes of the coefficients ($\bar{a_i}$ and $\bar{b_i}$ ) with the coupling parameter $\alpha$ for different rotation parameter $a$ are shown in Figs.(4)-(5).
It is shown that $\bar{a_i}$ and $\bar{b_i}$ depend on not only the black hole parameters and the coupling with Weyl tensor, but also the polarization direction and  the rotation direction of photon around black hole.
For PPM, we find that with the increase of $\alpha$ the coefficient $\bar{a}$ increases for different $a$.  For the retrograde photon for all $a$ and the prograde photon for the smaller $a$,  the coefficient $\bar{b}$ first decreases down to its minimum  with $\alpha$ and then increases up to its maximum with the further increase of $\alpha$; after that, it decreases with $\alpha$ again. For the prograde photon in the case with larger $a$, the unique difference is that as $\alpha<-0.06$, $\bar{b}$ first increases and then decreases with $\alpha$ rather than decreases monotonously.
The change of $\bar{a}$ with $a$ depends on the value of the coupling parameter $\alpha$. As $\alpha<0.06$,
the dependence of $\bar{a}$ for the retrograde photon on $a$ is similar to that in the case without Weyl correction in which  $\bar{a}$ increases with $a$. While for the larger $\alpha$, $\bar{a}$ decreases with the rotation parameter. For the prograde photon, the change of $\bar{a}$ with $a$ has also qualitative similar to that in the case without Weyl correction as $\alpha>-0.08$, but as $\alpha$ becomes more negative, $\bar{a}$ becomes a decreasing function of $a$.
The change of the coefficient $\bar{b}$ with $a$ also is related to the value of $\alpha$. For the retrograde photon, with increase of $\alpha$, the change of $\bar{b}$  with $a$ undergoes a process from increasing to decreasing and then to increasing, and finally to decreasing. For the prograde photon,  the change of $\bar{b}$ with $a$ undergoes only a process from increasing to decreasing.

For the PPL, we find that for the retrograde photon for all $a$ and the prograde photon for the smaller $a$, the coefficient $\bar{a}$  decreases monotonously with $\alpha$, but in the case with larger $a$ it first decreases and then increases and finally decreases quickly for the prograde photon.
The dependence of $\bar{b}$ with $\alpha$ becomes more complicated for PPL. In the smaller $|a|$ case,  $\bar{b}$ both for the retrograde and prograde photons first increases up to its maximum with $\alpha$ and then decreases down to its minimum with the further increase of $\alpha$, after that, it increases with $\alpha$ again. However, in the larger $|a|$ case, the peak in the curve $\bar{b}$ for a retrograde photon in the region $\alpha<0$ disappears gradually with increasing $|a|$, and finally the coefficient $\bar{b}$ as a whole becomes a function of increasing first and then decreasing. For the prograde photon, with increasing $|a|$, the trough in the middle region of the curve $\bar{b}$ becomes more deep, and the increasing monotonicity of $\bar{b}$ with $\alpha$ in the stronger coupling region (.i.e, $\alpha$ is larger) changes to decreasing. The variety of $\bar{a}$ with $a$ for PPL resembles that for PPM except for the prograde photon in the rapid rotation and strong coupling cases where $\bar{a}$ decreases with $a$ for PPL.
For the prograde photon, the variety of $\bar{b}$ with $a$ for PPL is converse to that for PPM. Thus, the Weyl coupling together with the rotation of black hole could result in the more complicated effects on strong gravitational lensing. Moreover, from Figs.(4) and (5), one can find that the coefficient $\bar{a}$ diverges as the coupling parameter $\alpha$ tends to the critical value $\alpha_{c1}$ or $\alpha_{c2}$, which implies that the deflection angle in the strong
deflection limit (\ref{alf1}) is not valid in the regime $\alpha<\alpha_{c1}$ for PPL and $\alpha>\alpha_{c2}$ for PPM, which is consistent with the previous discussion. Moreover, from Figs.(4) and (5), we also note that in the both cases of PPM and PPL, when $\alpha$ is close to the value corresponding to the turning point appeared in Figs.(2) and (3), there is accordingly a hump in the curve $\bar{a} (\alpha)$ or a sharp concave in the curve $\bar{b} (\alpha)$ for $a=0.4$. The main reason is that the coefficients
$\bar{a}$ and $\bar{b}$ (\ref{coa1}) are a function of the radius of the marginally circular photon orbit $x_{ps}$ and then their properties depend sharply on the behavior of $x_{ps}$. These similar features near the turning point may appear in the deflection angles and other observables in strong gravitational lensing. In Figs.(6) and (7), we plotted the change of the deflection angles
evaluated at $u=u_{ps}+ 0.003$ with $\alpha$ for PPL  and
PPM, respectively. Similarly, we find that the deflection angles in
the strong field limit have similar behaviors of the coefficient
$\bar{a}$ , which can be explained by a fact that the deflection angles are dominated by the logarithmic term in this limit.

\section{Effects of Weyl corrections on observables in strong gravitational lensing and time delay between relativistic images}

We are now in the position to estimate observable in strong gravitational lensing and time delay between relativistic images for
the coupled photon in a Kerr black hole spacetime. Since the source and observer actually are very far from the lensing galaxy,  the lens equation can be approximated well as \cite{Bozza3}
\begin{eqnarray}
\gamma=\frac{D_{OL}+D_{LS}}{D_{LS}}\theta-\alpha(\theta) \; \text{mod}
\;2\pi,
\end{eqnarray}
where $\gamma$ denotes the angle between the direction
of the source and the optical axis. $D_{LS}$ and $D_{OL}$ are
the lens-source distance and the observer-lens distance, respectively.
The angle $\theta=u/D_{OL}$ is the angular
separation between the lens and the image.
As in ref.\cite{Bozza3}, we here focus only on the simplest case where
the source, lens and observer are
highly aligned so that the angular distance of the $n-$th relativistic image from the lens can be simplified further as
\begin{eqnarray}
\theta_n\simeq\theta^0_n\bigg(1-\frac{u_{ps}e_n(D_{OL}+D_{LS})}{\bar{a}D_{OL}D_{LS}}\bigg),
\end{eqnarray}
with
\begin{eqnarray}
\theta^0_n=\frac{u_{ps}}{D_{OL}}(1+e_n),\;\;\;\;\;\;e_{n}=e^{\frac{\bar{b}+|\gamma|-2\pi
n}{\bar{a}}},\label{st1}
\end{eqnarray}
where the number $n$ is an integer and $\theta^0_n$ is the image position corresponding
to $\alpha=2n\pi$. In the limit $n\rightarrow \infty$, we find
$e_n\rightarrow 0$ and then the asymptotic position of a
set of images $\theta_{\infty}$ is related to the
minimum impact parameter $u_{ps}$ by a simpler form
\begin{eqnarray}
\theta_{\infty}=\frac{u_{ps}}{D_{OL}}.\label{ups}
\end{eqnarray}
To estimate further the coefficients $\bar{a}$ and $\bar{b}$, one needs at least another two observations, which could be obtained by separating the outermost image from all the others.
In a perfect situation, only the outermost relativistic image $\theta_1$ is separated as a single image and all the remaining ones are packed together at $\theta_{\infty}$ \cite{Bozza2,Bozza3}, and then  the angular separation $s$ and the relative magnitude $r_m$ between the images $\theta_1$  and $\theta_{\infty}$
can be simplified as \cite{Bozza2,Bozza3}
\begin{eqnarray}
s&=&\theta_1-\theta_{\infty}=
\theta_{\infty}e^{\frac{\bar{b}-2\pi}{\bar{a}}},\nonumber\\
r_m&=&2.5\log{\mathcal{R}_0}=2.5\log{\bigg(\frac{\mu_1}{\sum^{\infty}_{n=2}\mu_n}
\bigg)}
=\frac{5\pi}{\bar{a}}\log{e},\label{sR}
\end{eqnarray}
where $\mathcal{R}_0$ represents the ratio of the flux from the
first image $\theta_1$ and those from the all other images packed at $\theta_{\infty}$. Thus, we can estimate the coefficients $\bar{a}$, $\bar{b}$ and the minimum impact parameter $u_{ps}$ by measuring these three observations $s$, $r_m$,
and $\theta_{\infty}$ in the strong deflection limit. Comparing these data with the predicted values by the coupling theoretical model, we can identify the parameters of the central black hole in galaxies and examine further whether this coupling exists in the nature.

In our Milky Way Galaxy,  the mass of the central object is evaluated to be
$4.4\times 10^6M_{\odot}$ and its distance from the earth is around
$8.5kpc$ \cite{grf}, which means the ratio $GM/D_{OL}
\approx2.4734\times10^{-11}$ . With this data,  we can estimate numerically the values of the coefficients and observables in strong gravitational lensing by a Kerr black hole as the photon couples to Weyl tensor. In Figs.(8)-(13),
we plot the dependence of these observables on the
coupling constant $\alpha$ for different $a$, which are more complicated than those in Schwarzschild black hole spacetime. For the retrograde photon for all $a$ and the prograde photon for the smaller $a$,
we find that with the increase of the coupling constant $\alpha$, both the angular position of the relativistic
images $\theta_\infty$ and the relative magnitudes
$r_m$ decrease for PPM and increase for PPL. The variety of the angular
separation $s$ with $\alpha$ is converse to the varieties of
$\theta_\infty$ and $r_m$ with $\alpha$. For the prograde photon in the case with larger $a$, the unique difference is that $r_m$ for PPL first increases and then decreases and finally increases and the variety of $s$ with $\alpha$ is only converse to that of $r_m$ with $\alpha$.
The changes of these observables with $a$ also depend on the value of the coupling parameter $\alpha$. In the usual Kerr black hole spacetime without Weyl coupling, the observables $\theta_\infty$ and $r_m$
decreases with $a$, but $s$ increases. In the case of Weyl coupling, in general, the dependence of  three observables ($\theta_\infty$, $s$ ,$r_m$) on $a$ also obeys to those laws in the case without Weyl correction. However, in the case of the retrograde photon with the coupling parameter $\alpha\sim\alpha_{c1}$ or $\alpha\sim\alpha_{c2}$ the angular position $\theta_\infty$ is not a decreasing function of $a$ again. Moreover, in the case of the prograde photon with the stronger coupling (.i.e, $|\alpha|$ is larger ) $r_m$ increases rather than decreases with $a$. The presence of these different new features is not surprising because that the coupling between the photon and Weyl tensor changes the equation of motion
of the photon and makes the propagation of the light ray more complicated.
\begin{figure}
\begin{center}
\includegraphics[width=7cm]{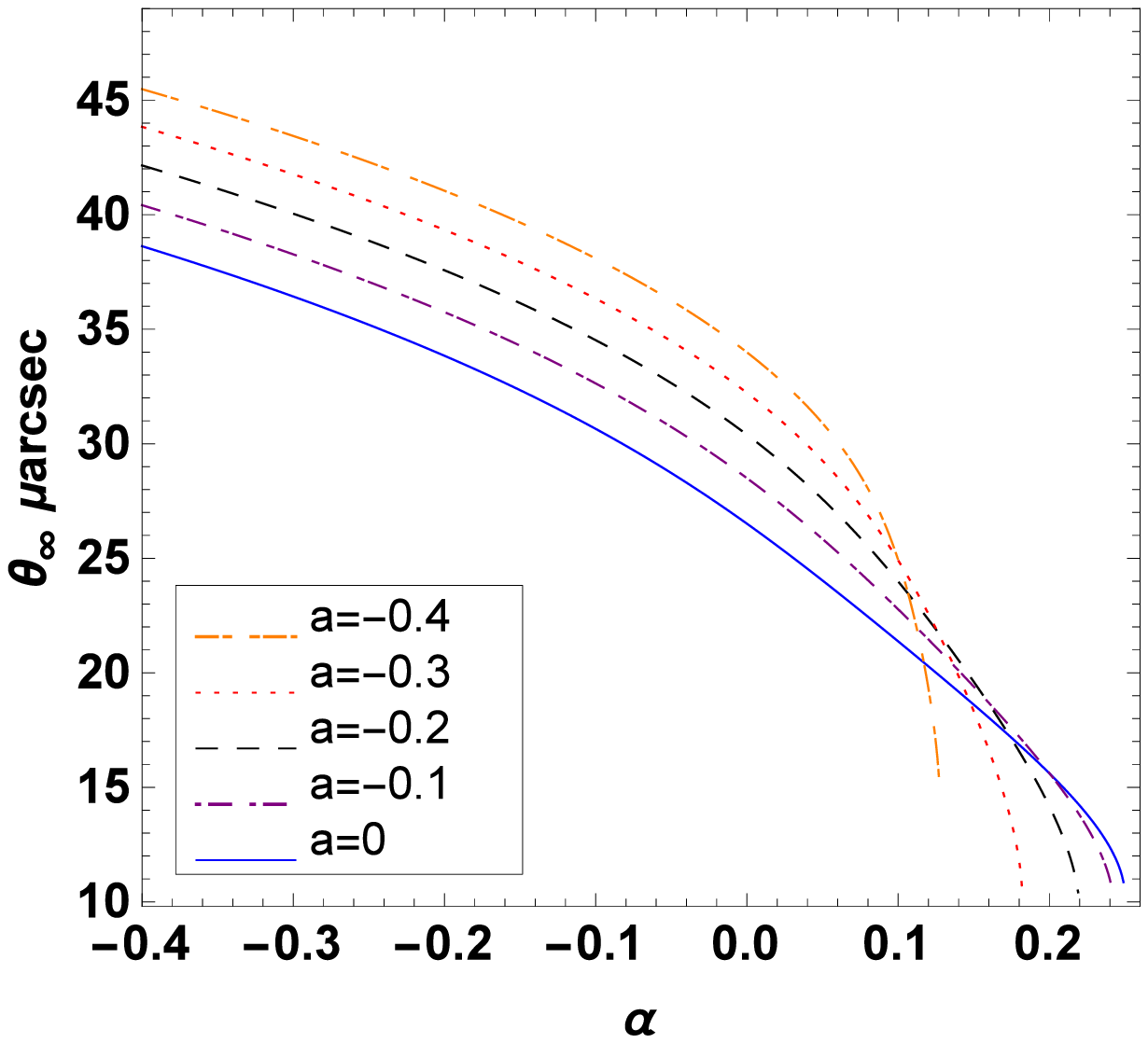}
\;\;\;\;\includegraphics[width=7cm]{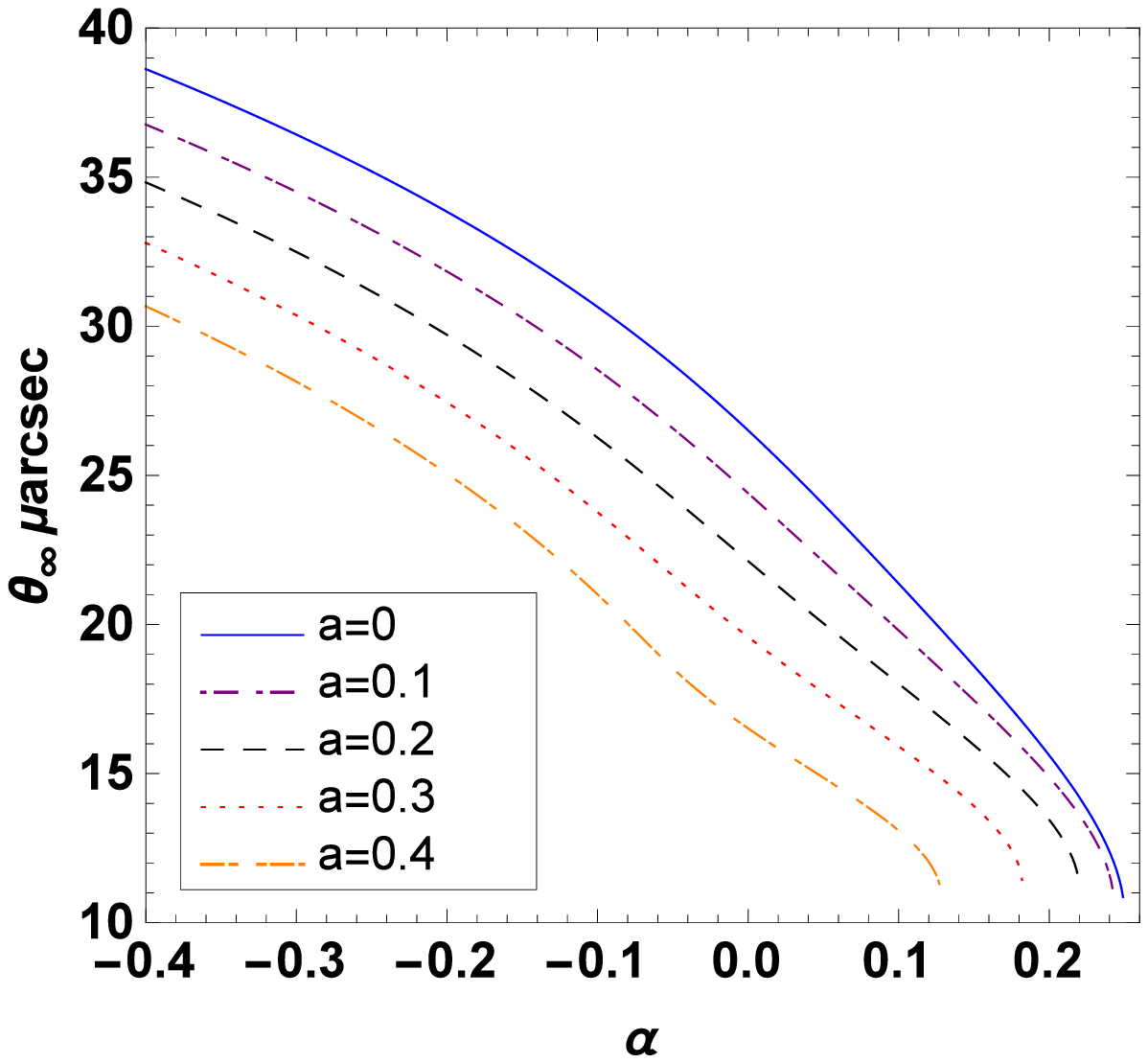}
\caption{Variety of the
innermost relativistic image $\theta_{\infty}$ for PPM with the coupling
parameter $\alpha$ for different $a$. Here, we set $2M=1$.}
\end{center}
\end{figure}
\begin{figure}
\begin{center}
\includegraphics[width=7cm]{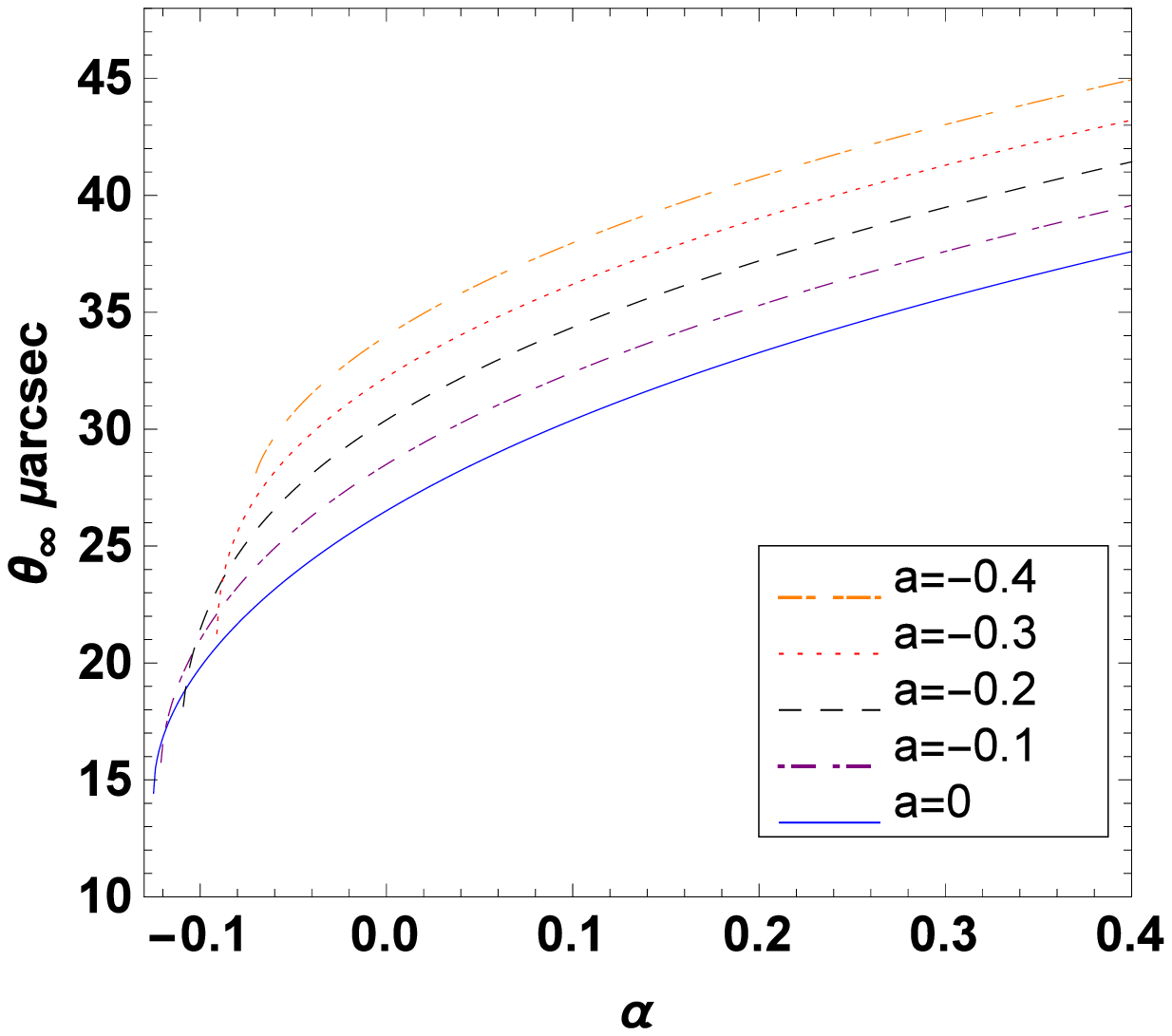}
\;\;\;\;\includegraphics[width=7cm]{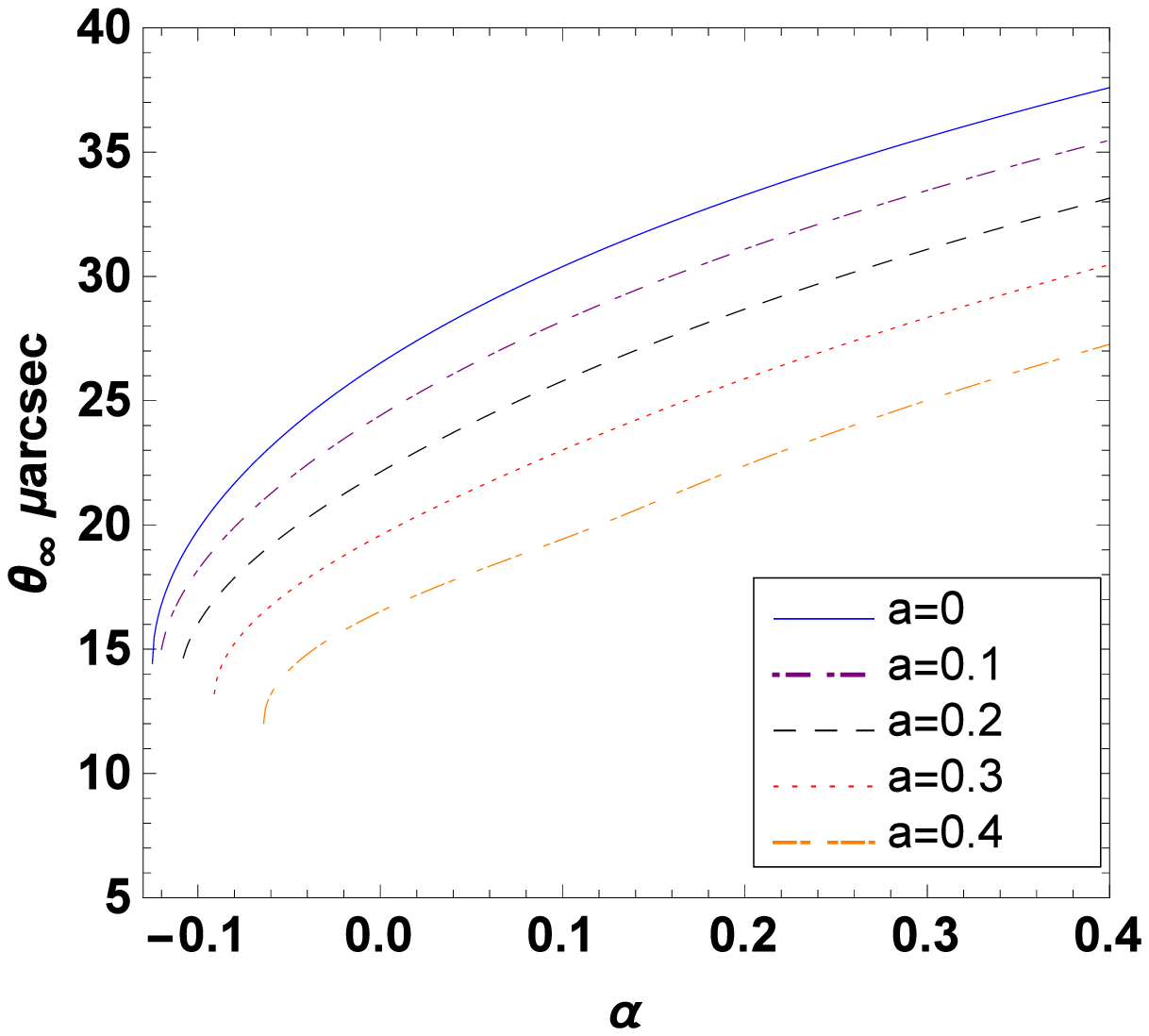}
\caption{Variety of the
innermost relativistic image $\theta_{\infty}$ for PPL with the coupling
parameter $\alpha$ for different $a$. Here, we set $2M=1$.}
\end{center}
\end{figure}
\begin{figure}
\begin{center}
\includegraphics[width=7cm]{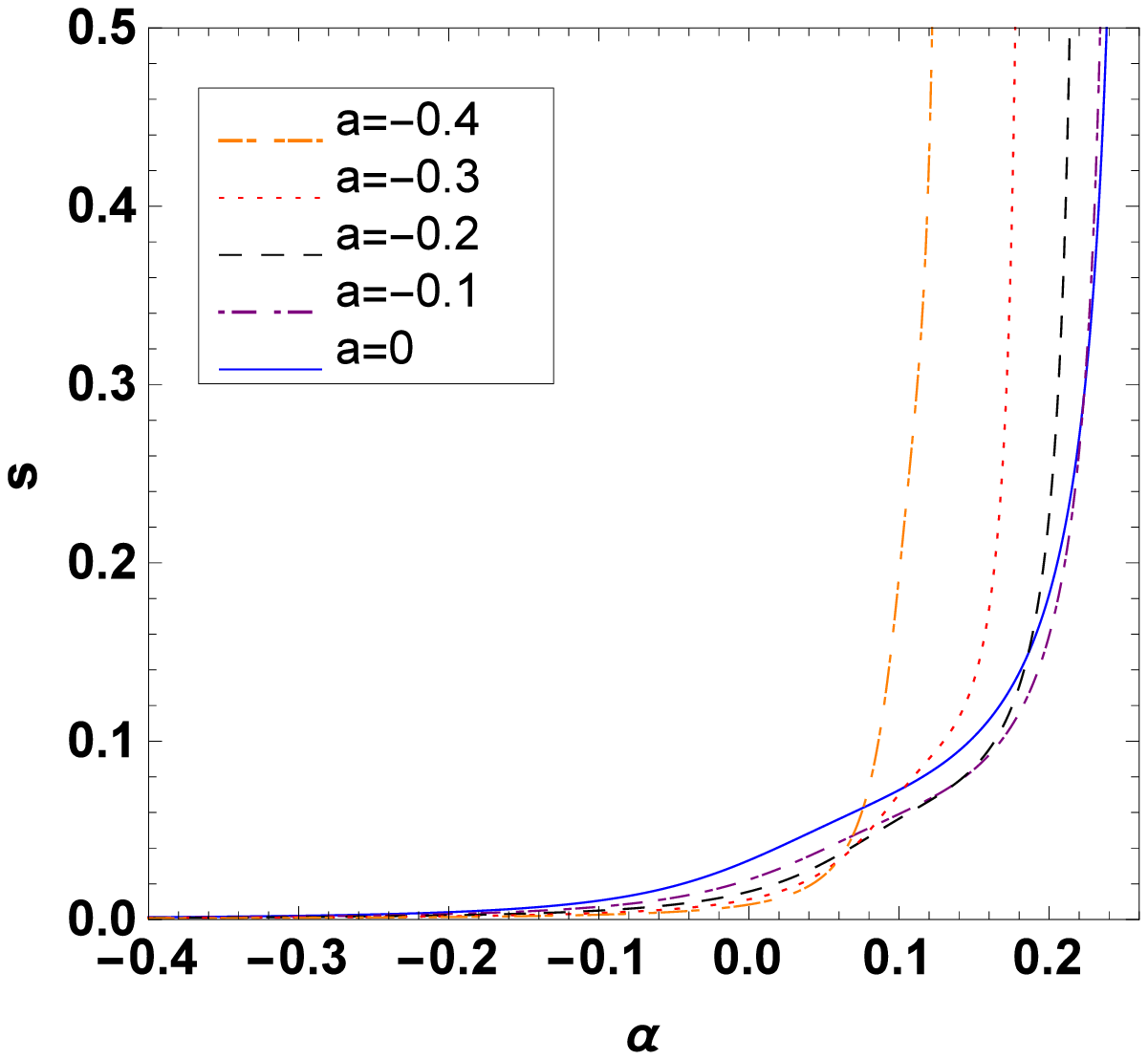}\;\;\;\;
\includegraphics[width=7cm]{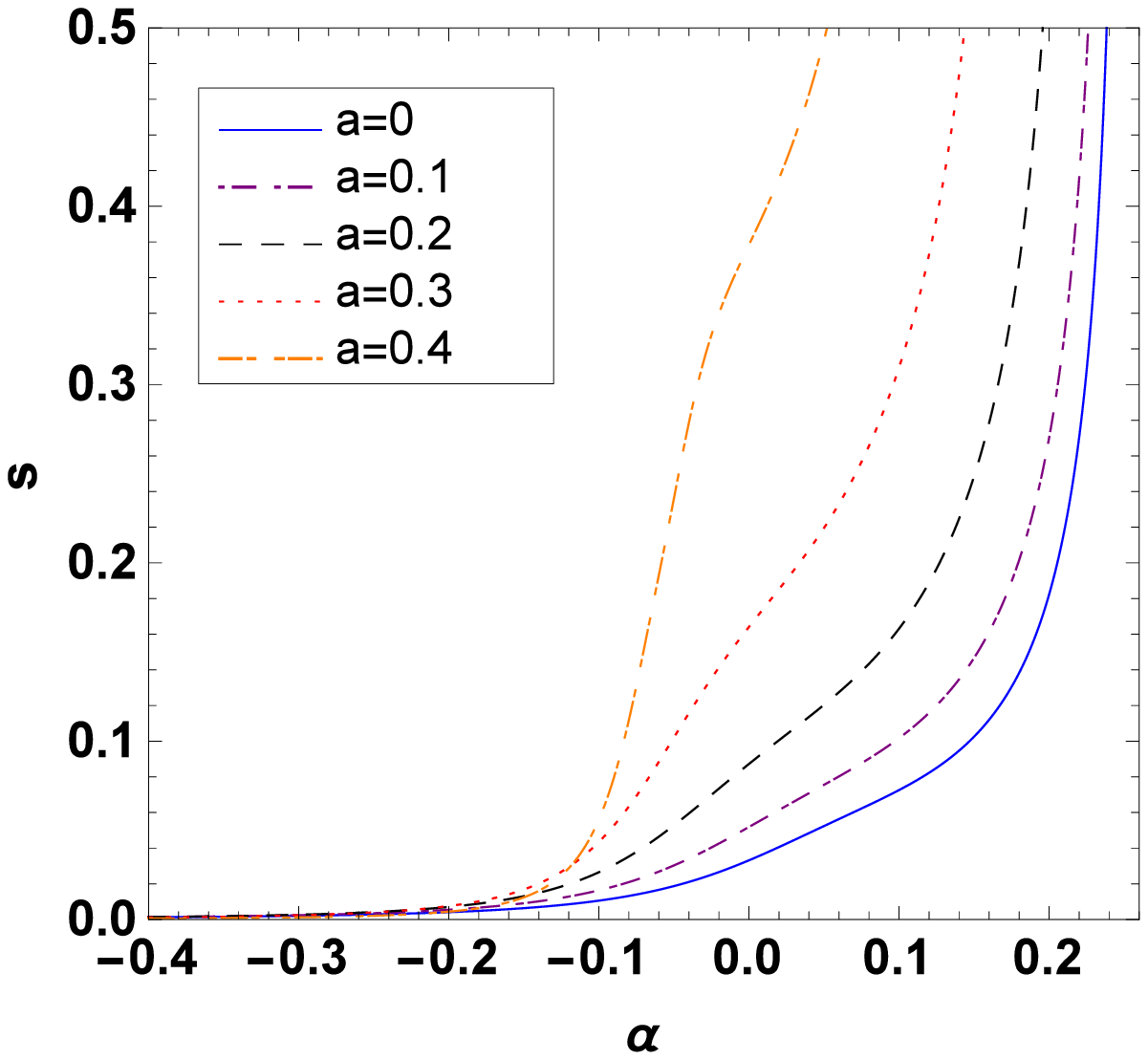}
\caption{Variety of the
angular separation $s$ for PPM with the coupling
parameter $\alpha$ for
different $a$. Here, we set $2M=1$.}
\end{center}
\end{figure}
\begin{figure}
\begin{center}
\includegraphics[width=7cm]{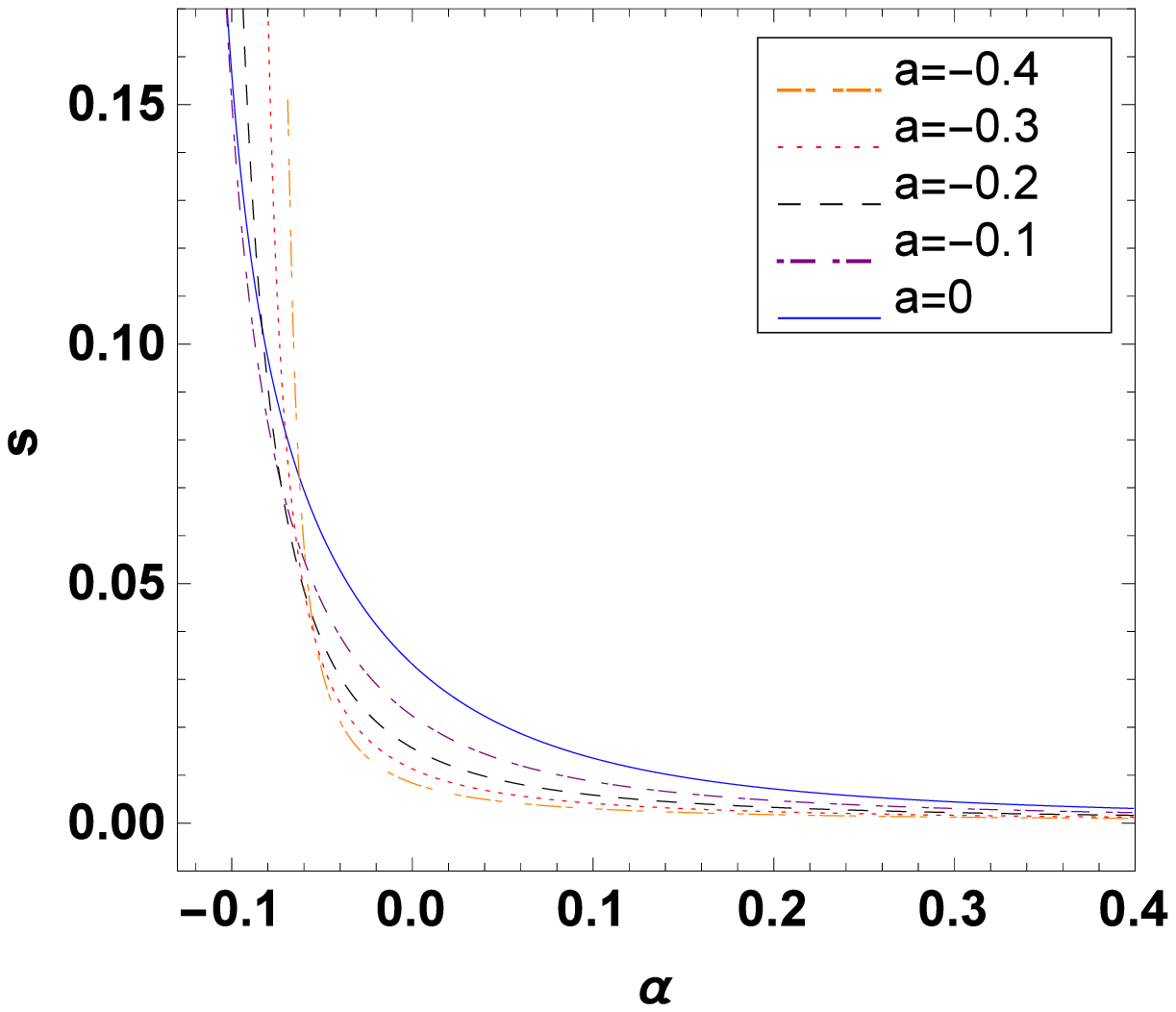}
\;\;\;\;\includegraphics[width=7cm]{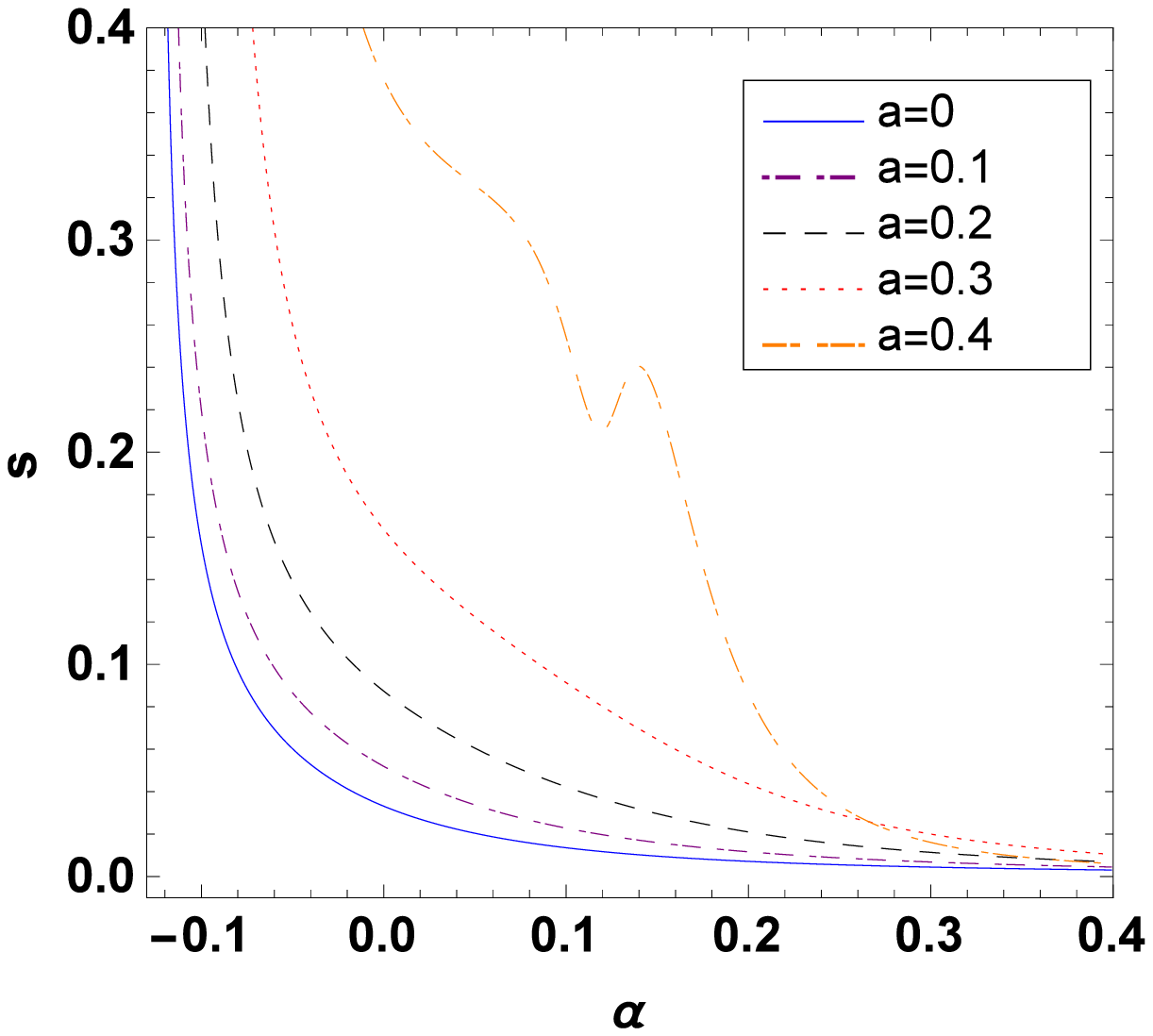}
\caption{Variety of the angular separation $s$ for PPL with the coupling
parameter $\alpha$ for different $a$. Here, we set $2M=1$.}
\end{center}
\end{figure}
\begin{figure}
\begin{center}
\includegraphics[width=7cm]{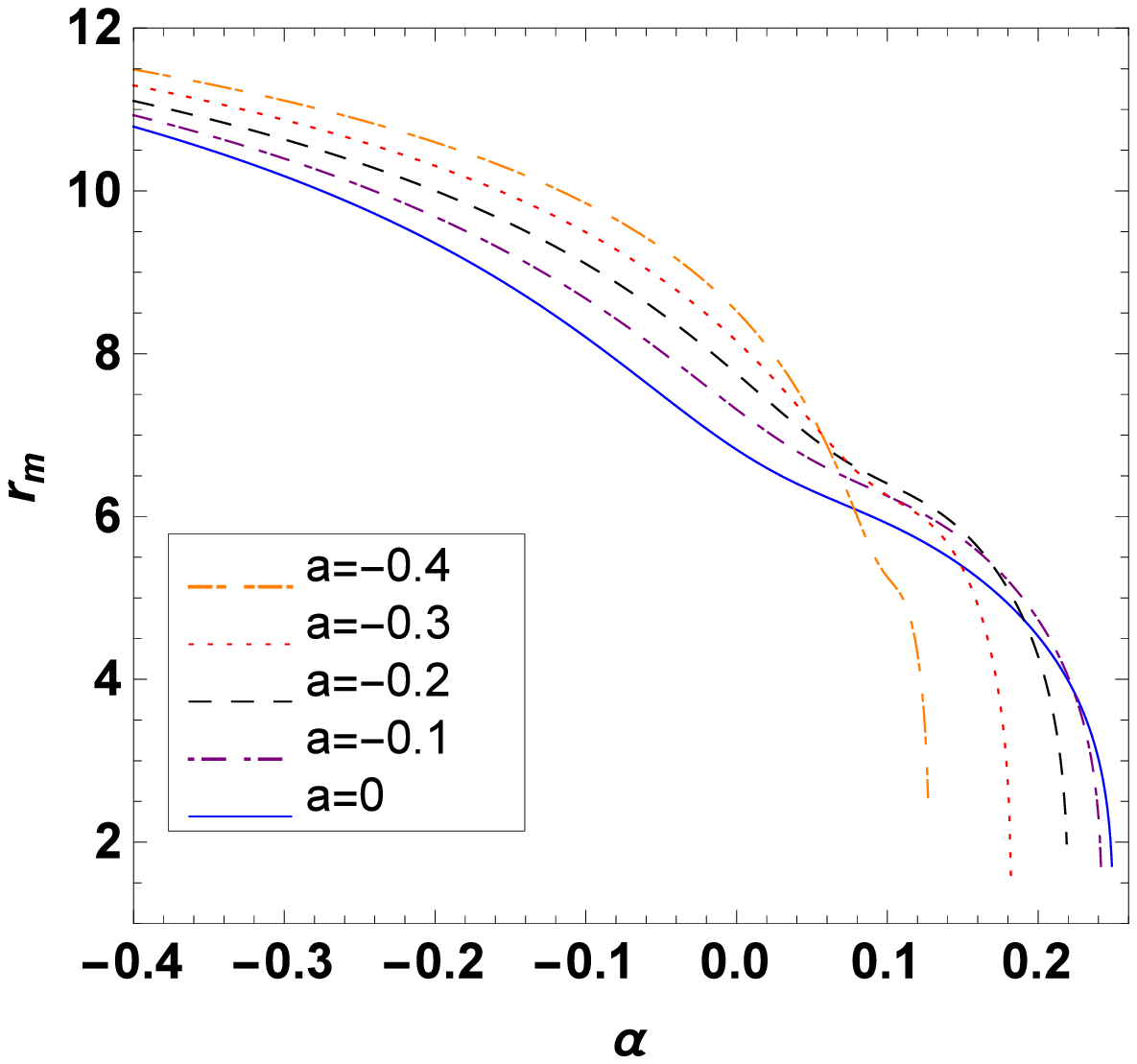}\;\;\;\;
\includegraphics[width=7cm]{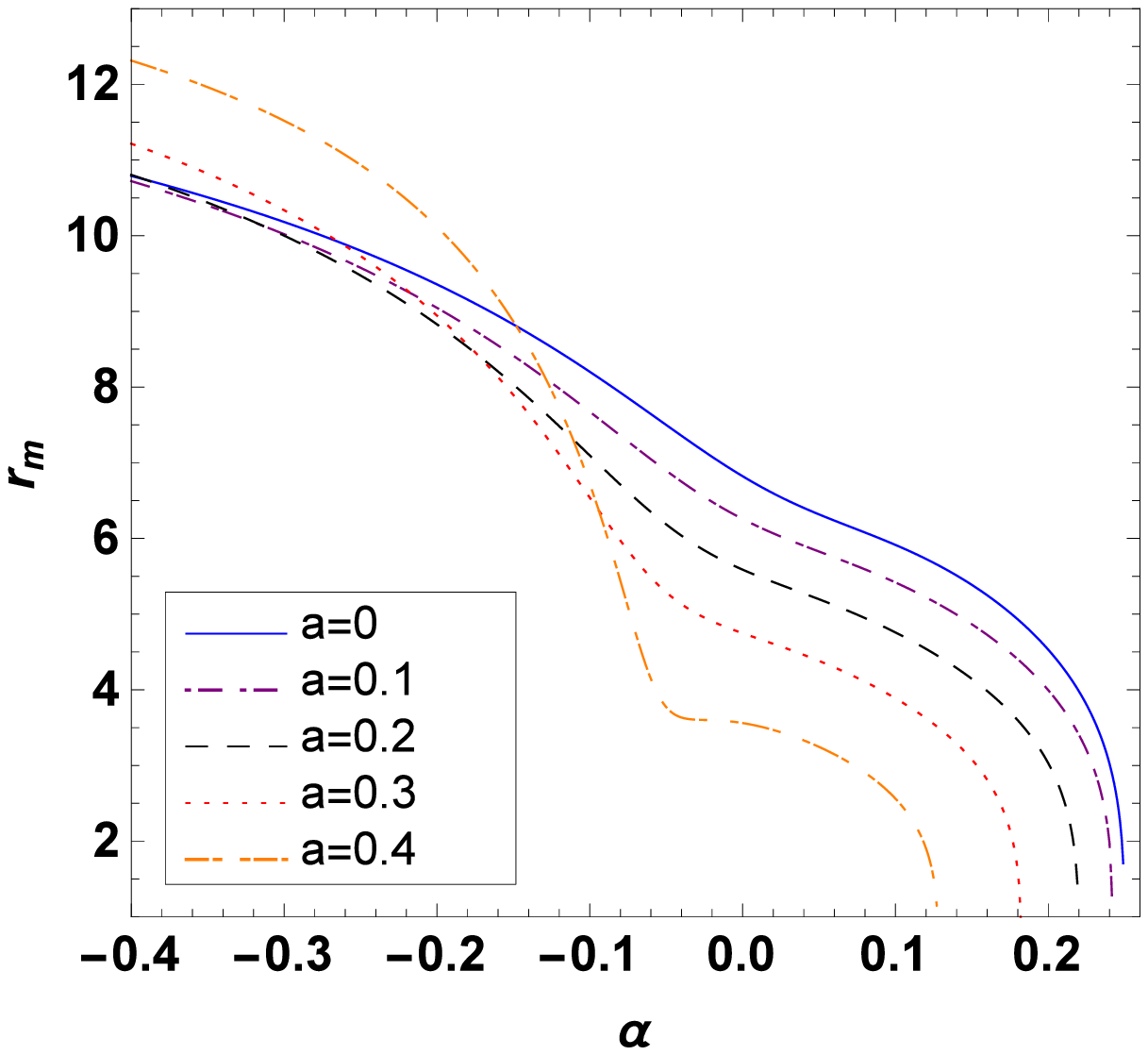}
\caption{Variety of the relative magnitudes $r_m$ for PPM with the coupling
parameter $\alpha$ for different $a$. Here, we set $2M=1$.}
\end{center}
\end{figure}
\begin{figure}
\begin{center}
\includegraphics[width=7cm]{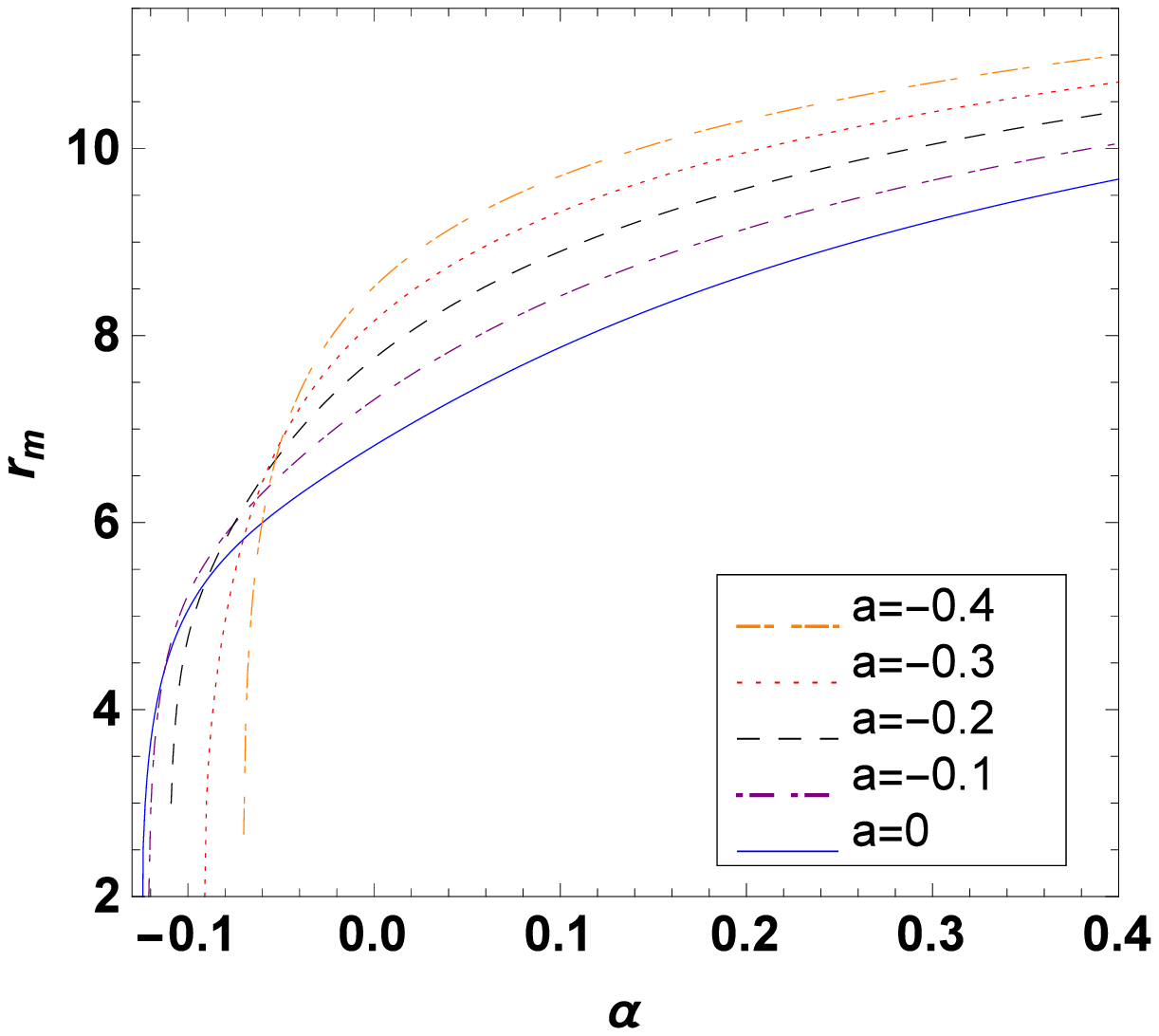}\;\;\;\;
\includegraphics[width=7cm]{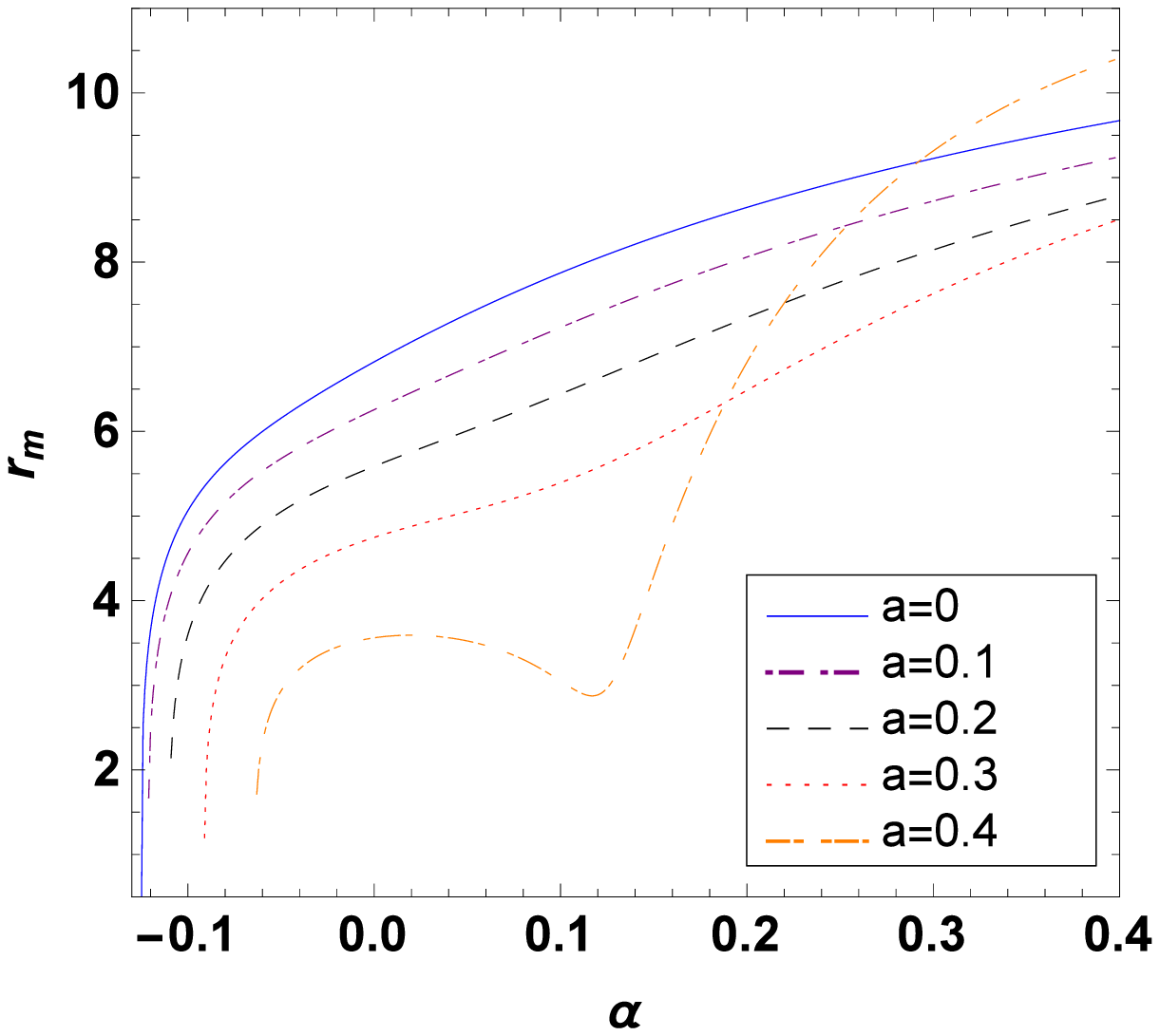}
\caption{Variety of the
relative magnitudes $r_m$ for PPL with the coupling parameter $\alpha$ for
different $a$. Here, we set $2M=1$.}
\end{center}
\end{figure}

The time delays between the relativistic images is
another important kind of observables in strong gravitational
lensing \cite{VBt,Tim2,Tim1,Yi1,Yi101}. From the null geodesics equation (\ref{cedi}), one has
\begin{eqnarray}
\frac{dt_i}{dx}&=&\frac{\sqrt{B_i(x)}[C_i(x)-J_iD_i(x)]}{
\sqrt{C_i(x)-2J_iD_i(x)-J_i^2A_i(x)}\sqrt{D_i(x)^2+A_i(x)C_i(x)}}.
\end{eqnarray}
The time for the PPM or PPL traveling from the source to the observer can be expressed as \cite{VBt}
\begin{eqnarray}
T_i(x_0)&=&2\int^{\infty}_{x_0}\bigg|\frac{dt_i}{dx}\bigg| dx-\int^{\infty}_{D_{OL}}\bigg|\frac{dt_i}{dx}\bigg| dx-\int^{\infty}_{D_{LS}}\bigg|\frac{dt_i}{dx}\bigg| dx.\label{T0}
\end{eqnarray}
The last two terms in Eq.(\ref{T0}) can be neglected actually since that observer and source are very far from
the black hole. Thus, the time delay between two photons travelling on different trajectories can be simplified further as
\begin{eqnarray}
T_{i1}-T_{i2}&=&2\int^{\infty}_{x_0,1}\bigg|\frac{dt_i}{dx}(x,x_{0,1})\bigg| dx-2\int^{\infty}_{x_{0,2}}\bigg|\frac{dt_i}{dx}(x,x_{0,2})\bigg| dx.\label{T10}
\end{eqnarray}
Assuming $x_{0,1}<x_{0,2}$, one has
\begin{eqnarray}
T_{i1}-T_{i2}&=&\tilde{T_i}(x_{0,1})-\tilde{T_i}(x_{0,2})+2\int^{x_{0,2}}_{x_{0,1}}
\frac{\tilde{P_i}(x,x_{0,1})}{\sqrt{A_{i0,1}}} dx+2\int^{\infty}_{x_{0,2}}\bigg[\frac{\tilde{P_i}(x,x_{0,1})}{\sqrt{A_{i0,1}}}-
\frac{\tilde{P_i}(x,x_{0,2})}{\sqrt{A_{i0,2}}}\bigg]dx,\label{T20}
\end{eqnarray}
with
\begin{eqnarray}
\tilde{P_i}(x,x_{0})&=&\frac{\sqrt{B_i(x)A_i(x_0)}[C_i(x)-J_iD_i(x)]}{\sqrt{C_i(x)}
\sqrt{D_i(x)^2+A_i(x)C_i(x)}},\nonumber\\
\tilde{T_i}(x_{0,1})&=&\int^{1}_{0}\tilde{R_i}(z,x_0)f(z,x_0)dz,\nonumber\\
\tilde{R_i}(z,x_0)&=&\frac{2x^2}{x_0\sqrt{C_i(z)}}\frac{\sqrt{B_i(z)|A_i(x_0)|}
[C_i(z)-J_iD_i(z)]}{\sqrt{D^2_i(z)+A_i(z)C_i(z)}}\bigg(1-\frac{1}{\sqrt{A_i(x_0)}f_i(z,x_0)}\bigg).
\end{eqnarray}
After a series of operations as in the calculation of the deflection, one can obtain finally\cite{VBt}
\begin{eqnarray}
\tilde{T_i}(u)=-\tilde{a_i}\log{\bigg(\frac{u}{u_{ps}}-1\bigg)}+\tilde{b_i}+\mathcal{O}(u-u_{ps}), \label{tf1}
\end{eqnarray}
with
\begin{eqnarray}
&\tilde{a_i}&=\frac{\tilde{R_i}(0,x_{ps})}{\sqrt{q_i(x_{ps})}}, \nonumber\\
&\tilde{b_i}&= -\pi+\tilde{b}_{iR}+\tilde{a_i}\log{\bigg\{\frac{2q_i(x_{ps})C_i(x_{ps})}
{u_{ips}A_i(x_{ps})[D_i(x_{ps})+J_iA_i(x_{ps})]}\bigg\}}, \nonumber\\
&b_{iR}&=\int^{1}_{0}[\tilde{R_i}(z,x_{ps})f_i(z,x_{ps})-\tilde{R_i}(0,x_{ps})f_{i0}(z,x_{ps})]dz,\label{cot1}
\end{eqnarray}
which diverges logarithmically as the deflection angle in the strong-field limit. Assuming the source, the lens and the observer are aligned almost in a line, the time delay between a $n$-loop and a $m$-loop relativistic image can be approximated as
\begin{eqnarray}
\Delta T_{in,m}=\Delta T^0_{in,m}+\Delta T^1_{in,m}, \label{tfs1}
\end{eqnarray}
where
\begin{eqnarray}
\Delta T^0_{in,m}&=&2\pi(n-m)\frac{\tilde{a_i}}{\bar{a_i}},\nonumber\\
\Delta T^1_{in,m}&=&2\sqrt{\frac{B_i(x_{ps})}{A_i(x_{ps})}}
\sqrt{\frac{x^2_{ps}u_{ips}A_i(x_{ps})[D_i(x_{ps})+J_iA_i(x_{ps})]}{
2q_i(x_{ps})C_i(x_{ps})}}e^{\frac{\bar{b_i}}{2\bar{a_i}}}
\bigg[e^{-\frac{m\pi}{\bar{a_i}}}-e^{-\frac{n\pi}{\bar{a_i}}}\bigg]. \label{tfs2}
\end{eqnarray}
In Figs.(14) and (15), we present the time delay between the first relativistic image and the second one for PPM and PPL in a Kerr black hole spacetime, respectively. It is shown that the time delay $\Delta T_{2,1}$ decreases with the rotation parameter $a$ except in the case with $a<0$ and the value $\alpha$ near the critical value $\alpha_{c1}$ or $\alpha_{c2}$ in which $\Delta T_{2,1}$ is an increasing function of $a$.
With increase of the coupling parameter $\alpha$, it decreases for PPM and increases for PPL.
\begin{figure}
\begin{center}
\includegraphics[width=7cm]{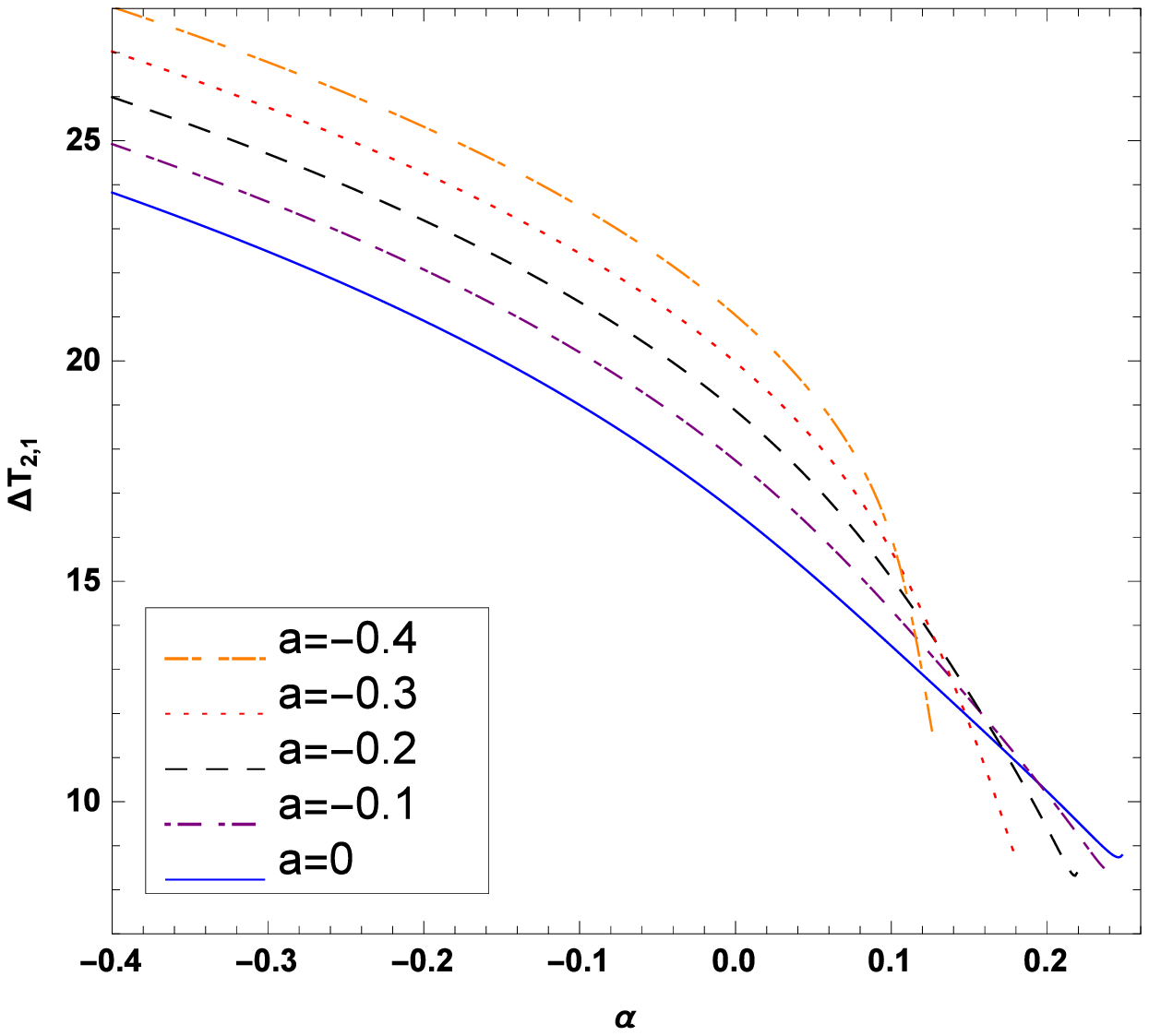}\;\;\;\;
\includegraphics[width=7cm]{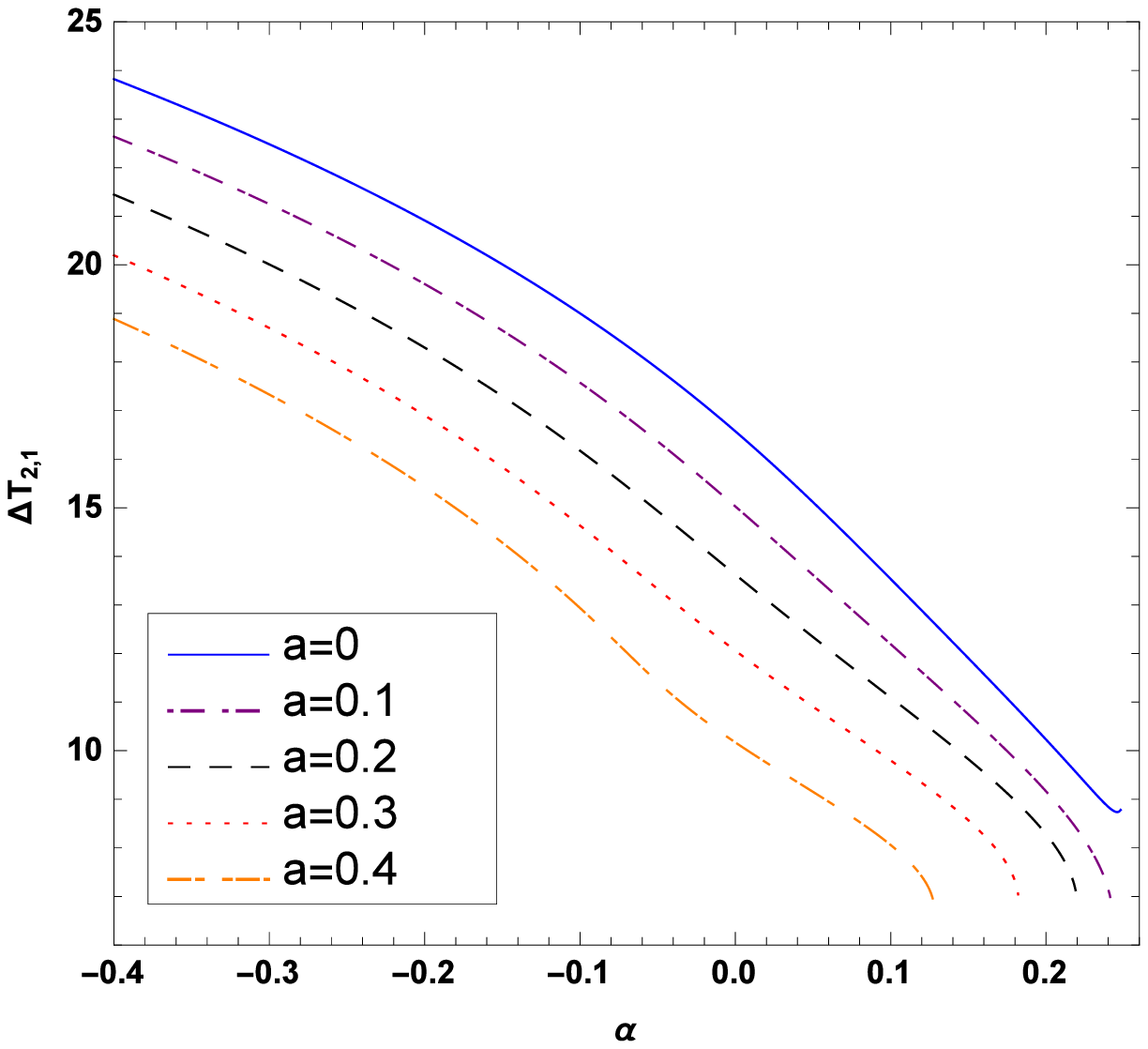}
\caption{Variety of the time delays between the relativistic images for PPM with the coupling parameter $\alpha$ for different $a$ in a Kerr black hole spacetime. Here, we set $2M=1$, $n=2$ and $m=1$.}
\end{center}
\end{figure}
\begin{figure}
\begin{center}
\includegraphics[width=7cm]{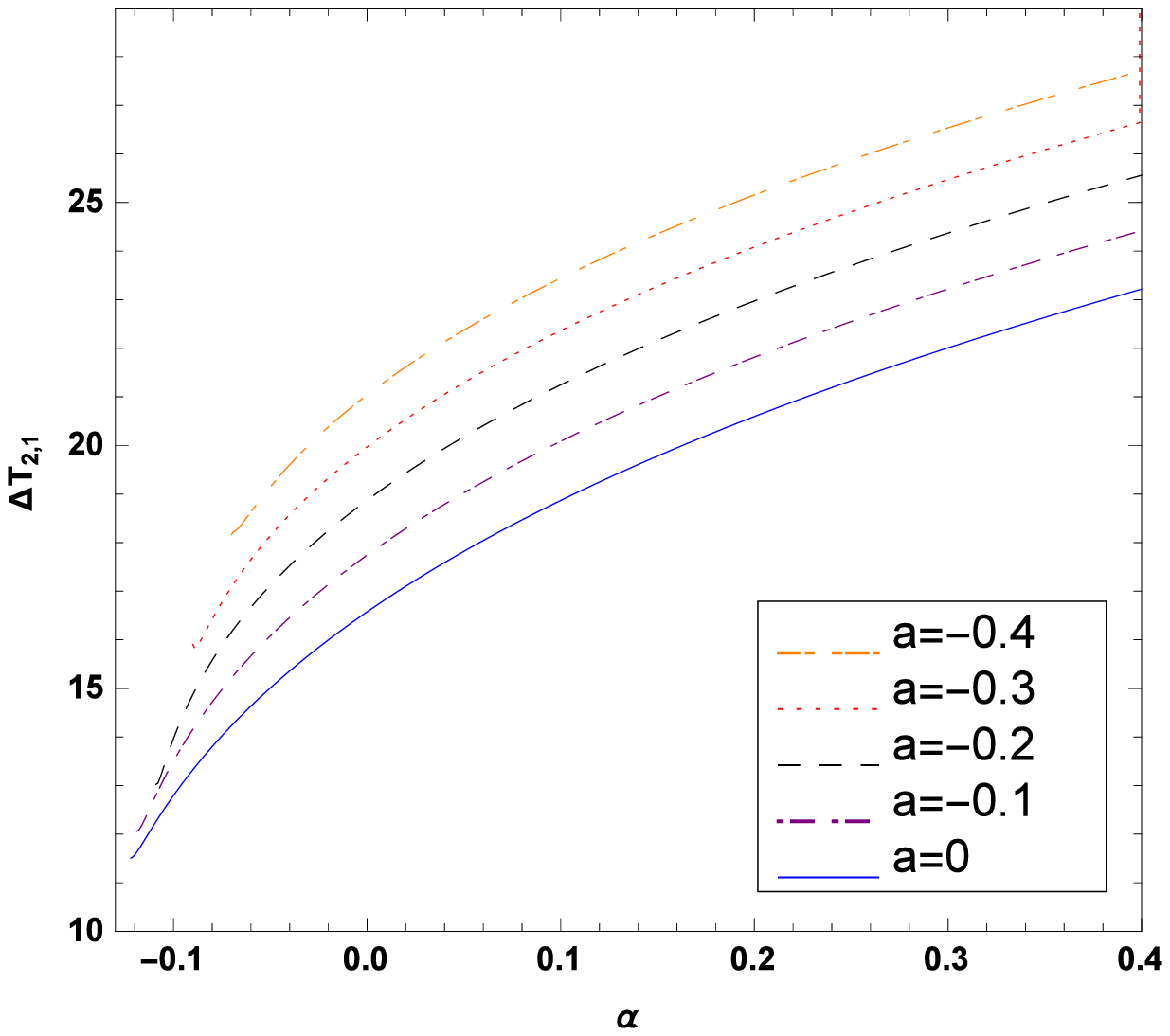}\;\;\;\;
\includegraphics[width=7cm]{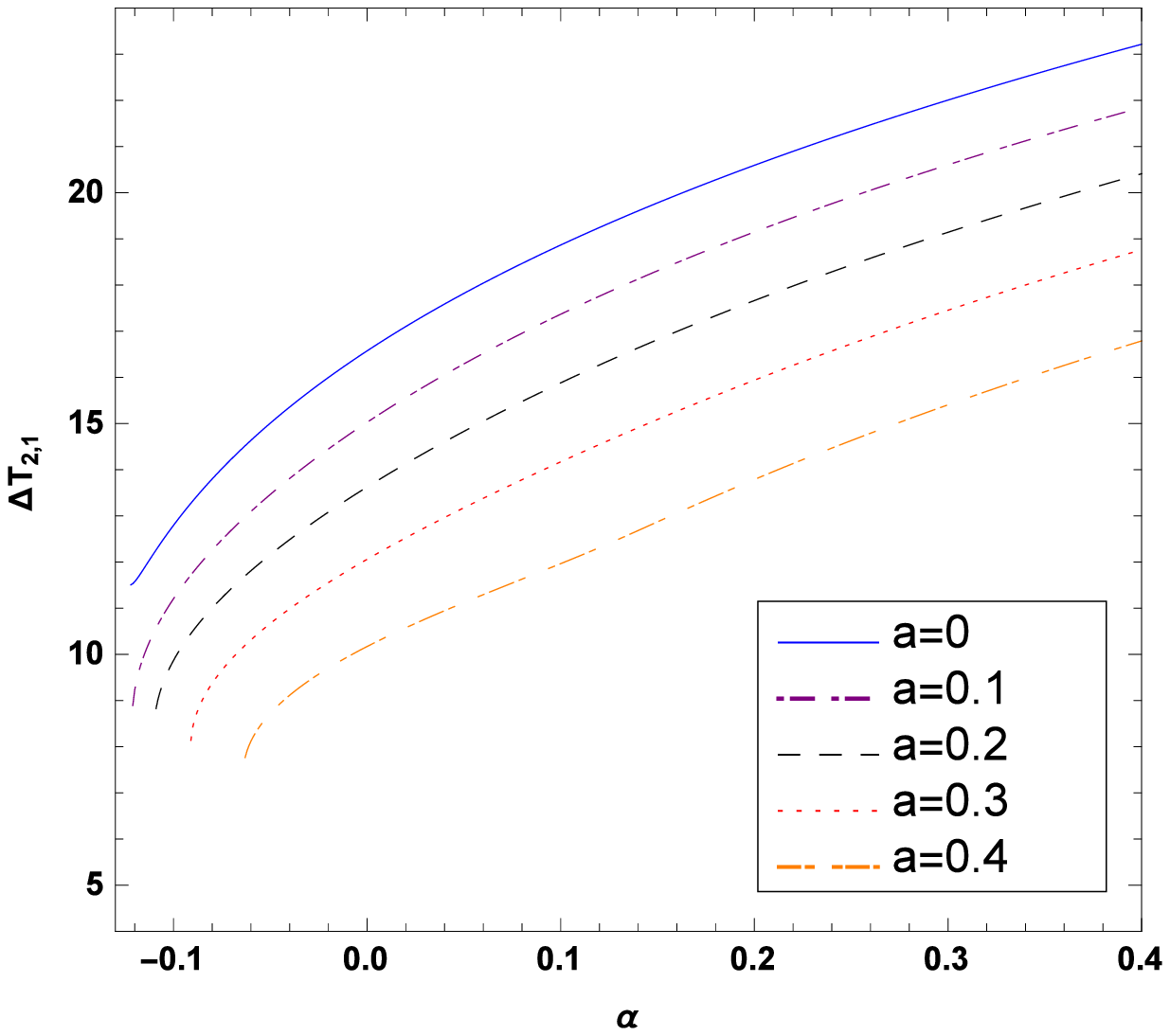}
\caption{Variety of the time delays between the relativistic images for PPL with the coupling parameter $\alpha$ for different $a$ in a Kerr black hole spacetime. Here, we set $2M=1$, $n=2$ and $m=1$.}
\end{center}
\end{figure}

\section{summary}

In this paper, we have investigated firstly the equation of motion for the
photon coupled to Weyl tensor in a Kerr black hole spacetime and then probed further the corresponding strong
gravitational lensing.  The coupling constant and the polarization
direction of photon together with the rotation parameter of black hole make  the propagation of the coupled photons more complicated, which brings about some new features for the physical quantities including the marginally circular photon orbit, the deflection angle, the observational gravitational lensing variables and the time delay between two relativistic images. The marginally circular photon orbit outside the event horizon exists only in the regime $\alpha< \alpha_{c1}$ for PPM and in the regime $\alpha> \alpha_{c2}$ for PPL. The critical values $\alpha_{c1}$ decreases with $|a|$ and $\alpha_{c2}$ increases  with $|a|$. The main new behavior of $x_{ps}$ originating from the rotation parameter is that in the case with the larger $a$ the radius $x_{ps}$ for PPL with $a>0$ first increases with $\alpha$ and then decrease  and finally increases again, while in the non-rotating case, $x_{ps}$ for PPL increases  monotonously with the coupling parameter. The new effects on $x_{ps}$ originating from the coupling parameter appear in the region with $\alpha$ near the critical value in which $x_{ps}$ for the retrograde photon increases with $a$. It modifies a common feature of $x_{ps}$ in a rotating black hole spacetime since $x_{ps}$  always decreases monotonously with $a$ in the case without Weyl coupling. These new effects make the changes of coefficients $\bar{a}$ and $\bar{b}$ with $a$ and $\alpha$ become more complicated.

Combining with the supermassive central object in our
Galaxy, we estimated the observables including time delays between the relativistic images in the strong gravitational
lensing for the photons coupled to Weyl tensor. The dependence of these observables on the coupling and the rotation parameter are different for two different kinds of coupled photons with different polarizations.
Due to the presence of the rotation parameter, we note that the observable  $r_m$ ($s$ ) for PPL with $a>0$ does not increases (decreases) monotonous with $\alpha$ in the case with larger $a$, which is different from those in Schwarzschild black hole spacetime. The Weyl coupling also change the relationship between the observable ($\theta_\infty$, $s$, $r_m$, $\Delta T_{2,1}$) and the rotation parameter $a$ for the retrograde photon as the coupling parameter is near the critical value $\alpha_{c1}$ or $\alpha_{c2}$.  Moreover, in the case of the prograde photon with the stronger coupling the relative magnitude $r_m$ increases rather than decreases with $a$. The main reason for the emergence of these different new features in strong gravitational lensing is  that the coupling between the photon and Weyl tensor changes the equation of motion of the photon and makes the propagation of the light ray more complicated.
It would be of interest to generalize our study to other rotating black hole
spacetimes.

\section{\bf Acknowledgments}

This work was  partially supported by the National Natural Science
Foundation of China under Grant No.11275065, the construct
program of key disciplines in Hunan Province and the Open Project Program of State Key Laboratory of Theoretical Physics, Institute of Theoretical Physics, Chinese Academy of Sciences, China (No.Y5KF161CJ1). J. Jing's work was
partially supported by the National Natural Science Foundation of
China under Grant No.11475061.

\end{document}